\documentclass{aa}
\usepackage[flushleft]{threeparttable}
\usepackage{graphicx}

\newcommand{\SNR}{\mbox{signal-to-noise}}
\newcommand{\etal}{{et al.}}
\newcommand{\nancay}{Nan\c{c}ay}

\newcommand{\TorchiSA}{Torchinsky,~S.A.}
\newcommand{\enancay}{\mbox{EMBRACE@\nancay}}

\newcommand{\degrees}{$^\circ$}
\newcommand{\ie}{{i.e.}}
\newcommand{\microsec}{\mbox{~$\mu$s}}

\newcommand{\printcomment}[1]{
  \vspace*{3ex}
  \begin{center}
    \fbox{
    \begin{minipage}{0.7\linewidth}
      {\large\it\bf #1}
    \end{minipage}
    }
  \end{center}
  \vspace*{3ex}
}
\renewcommand{\printcomment}[1]{}
\newcommand{\refereedelete}[1]{}
\newcommand{\dontuse}[1]{}
\newcommand{\modified}[1]{{#1}}
\newcommand{\modifiedmath}[1]{{#1}}

\begin{document}

\title{Characterization of a dense aperture array for~radio~astronomy}

\author{S.A.~Torchinsky\inst{1}
\and A.O.H.~Olofsson\inst{2}
\and B.~Censier\inst{1}
\and A.~Karastergiou\inst{3,4,5}
\and M.~Serylak\inst{5,6}
\and P.~Picard\inst{1}
\and P.~Renaud\inst{1}
\and C.~Taffoureau\inst{1}
}

\institute{
Station de radioastronomie de Nan\c{c}ay, Observatoire de Paris, CNRS, France, torchinsky@obs-nancay.fr
\and Onsala Space Observatory, Chalmers University of Technology, Gothenburg, Sweden
\and Astrophysics, University of Oxford, Denys Wilkinson Building, Keble Road, Oxford OX1 3RH, UK
\and Department of Physics and Electronics, Rhodes University, PO Box 94, Grahamstown 6140, South Africa
\and Department of Physics \& Astronomy, University of the Western Cape, Cape Town, South Africa
\and SKA South Africa, Cape Town, South Africa
}

\abstract{
\enancay\ is a prototype instrument consisting of an array of 4608
densely packed antenna elements creating a fully sampled, unblocked
aperture.  This technology is proposed for the Square Kilometre Array
and has the potential of providing an extremely large field of view
making it the ideal survey instrument.  We describe the system,
calibration procedures, and results from the prototype.
}


\titlerunning{EMBRACE@Nan\c{c}ay}
\authorrunning{Torchinsky et al.}
\maketitle

\section{Introduction}

\noindent The Square Kilometre Array (SKA) \citep{dewdneyska} will be
the largest radio astronomy facility ever built with more than
ten~times the equivalent collecting area of currently available
facilities. The SKA will primarily be a survey instrument with
exquisite sensitivity and an extensive field of view, providing an
unprecedented mapping speed. This capability will enormously advance
our understanding of fundamental physics including gravitation, the
formation of the first stars, and the origin of magnetic fields, and it
will give us a new look at the variability of the Universe with a
survey of transient phenomena.

\begin{figure}[ht!]
 \centering
 \includegraphics[width=0.95\linewidth,clip]{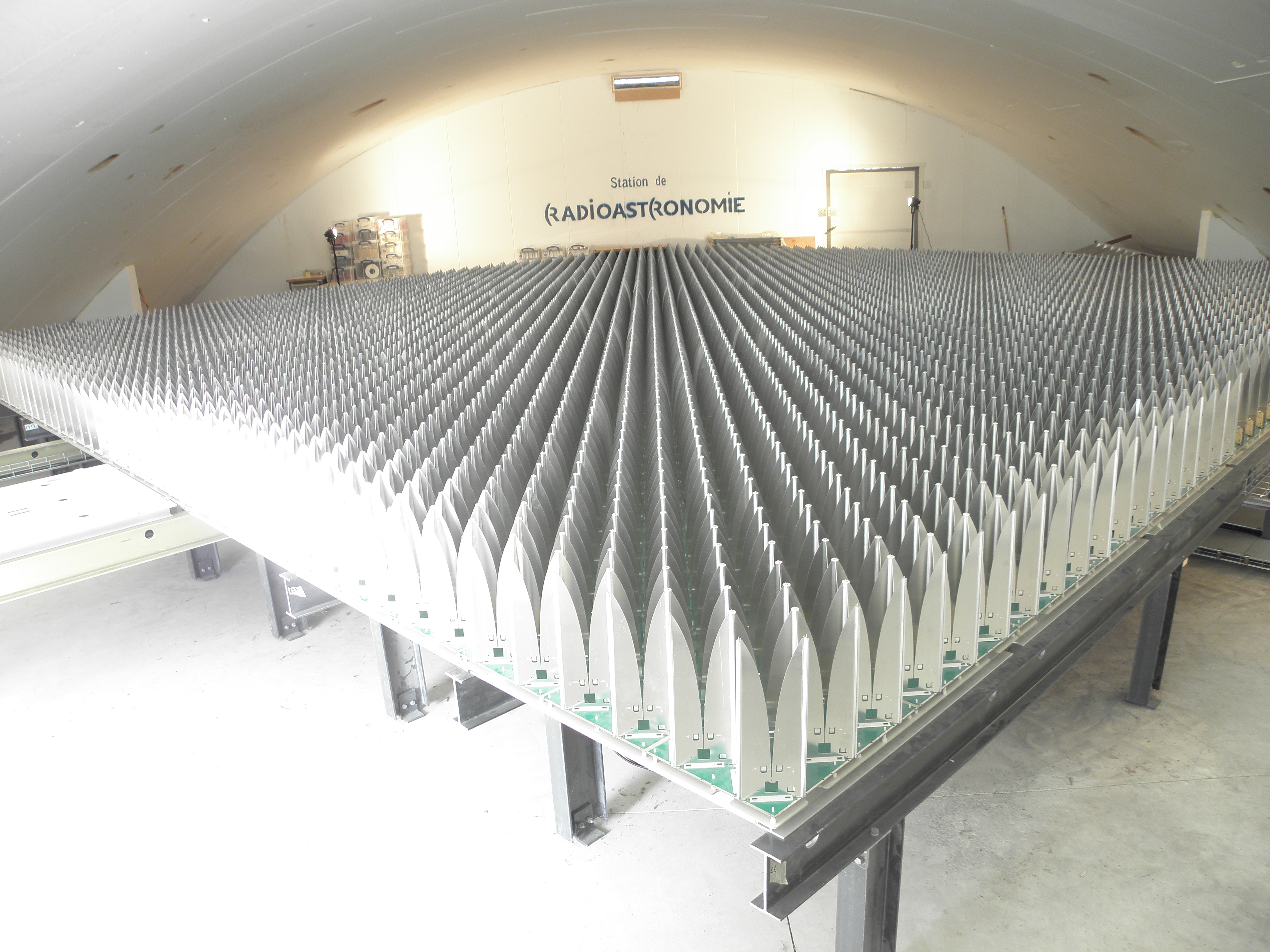}
  \caption{\enancay\ is composed of 4608 Vivaldi antenna elements separated from each other by 12.5~cm, making it a dense array for frequencies above 1200~MHz.  \enancay\ measures 8.42~m~$\times$~8.42~m for a total area of 70.8~m$^2$. }
  \label{fig:embrace}
\end{figure}

A revolution in radio-receiving technology is underway with the
development of densely packed phased arrays.  This technology can
provide an exceptionally large field of view, while at the same time
sampling the sky with high angular resolution.  The \nancay\ radio
observatory is a major partner in the development of dense phased
arrays for radio astronomy, working closely with The Netherlands
Institute for Radio Astronomy (ASTRON).  The joint project is called
EMBRACE (Electronic MultiBeam Radio Astronomy Concept).  Two EMBRACE
prototypes have been built.  One at Westerbork in The Netherlands (called EMBRACE@Westerbork) and
one at \nancay\ (\enancay, see Fig.~\ref{fig:embrace}).  The EMBRACE
prototypes are recognized as ``Pathfinders'' for the SKA project.
Conclusions from the EMBRACE testing will directly feed into the SKA
and will have a decisive impact on whether dense array
technology is used for the SKA.

The date for selecting technology for the SKA is 2018.  If dense
arrays are not selected for the SKA, then the SKA will have a much
reduced mapping speed compared to what has come to be expected by the
astronomical community.  It is therefore crucial that work on EMBRACE
succeeds in showing the viability of dense arrays for radio astronomy.

The two EMBRACE stations began with an initial period of engineering
testing on the partially complete arrays
\citep{olofsson_limelette,wijnholds_limelette}.
\enancay\ has been fully operational since 2011
and now performs regularly scheduled
astronomical observations, such as pulsar observations and
extragalactic spectral line observing~\citep{2013sf2a.conf..439T, 2015JInst..10C7002T}.
EMBRACE system characteristics, such as beam main lobe and system
temperature, are behaving as expected.  EMBRACE has long-term
stability, and after four years of operation it continues to prove itself
as a robust and reliable system capable of sophisticated radio
astronomy observations.

\section{EMBRACE system description}
\label{sec:sysdesc}
\enancay\ is a phased-array of 4608 densely packed antenna elements
(64 tiles of 72 elements each).  For mechanical and electromagnetic
performance reasons, \enancay\ has, in fact, 9216 antenna elements,
but only one polarization (4608 elements) has fully populated signal
chains.  The orientation of the linear polarized elements at
\nancay\ is with the electric field sensitivity in the north-south
direction.  The tuning range for EMBRACE was originally designed to be
from 500~MHz to 1500~MHz, however, the extremely powerful digital
television transmitter in UHF channel 66 at 834~MHz created problems
of saturation and intermodulation of the analogue components,
especially the beamformer chip.  As a result, high-pass filters were
added to the system to restrict the tuning range to frequencies above
900~MHz.  For more details on EMBRACE architecture, see
\citet{kant_limelette, kant11}

\subsection{Hierarchical analogue beam forming}

\begin{figure*}[ht!]
 \centering \includegraphics[width=0.9\textwidth,clip]{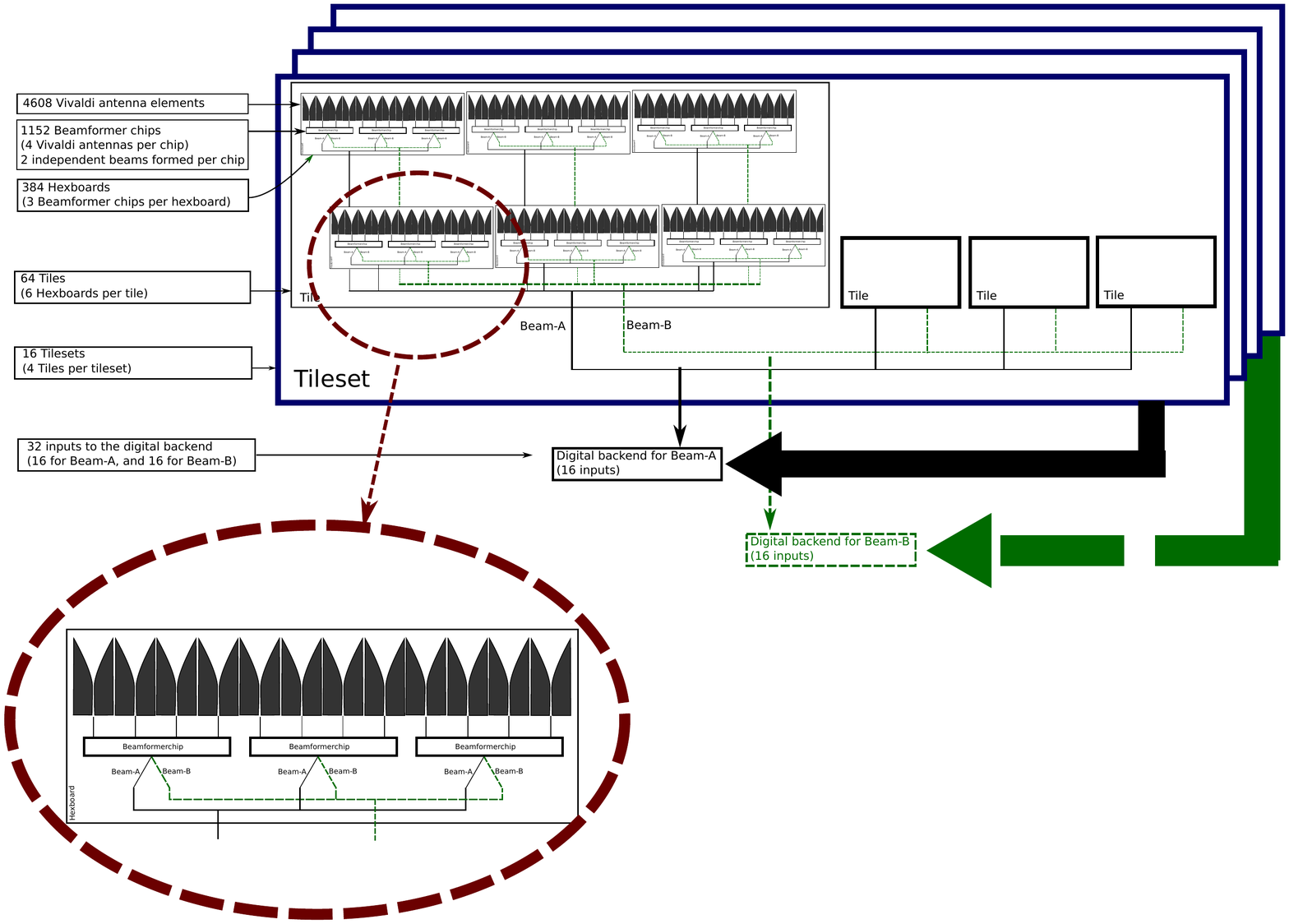}
 \caption{\enancay\
 system architecture consists of a four level hierarchy of analogue
 beamforming starting with 4608 Vivaldi antenna elements leading
 finally to 16 inputs to each of two digital
 backends.}  \label{fig:architecture}
\end{figure*}

\enancay\ uses a hierarchy of four levels of analogue beamforming
leading to 16~inputs to the LOFAR backend system used for digital
beamforming~\citep{2013A&A...556A...2V}.  An overview of the hierarchy
is shown in the frontend architecture diagram
(Fig.~\ref{fig:architecture}) and Table~\ref{table:components} gives
a summary of the total number of components at each stage of analogue
beamforming.

The first level of beamforming is done for four Vivaldi elements
within the integrated circuit ``beamformer chip'' developed at
\nancay~\citep{bosse10}.  This chip applies the phase shifts necessary
to four antenna elements to achieve pointing in the desired direction.
The phase shift required for each Vivaldi element is calculated from
the array geometry of a tileset (4~tiles) for a given pointing
direction.  In this way, the subsequent stages of analogue summing are
done by simple combining.  No further phase shifts are applied.

The beamformer chip forms two independent beams for each set of four
antenna elements.  The beamformer chip splits the analogue signal from
the antennas into two signals and then applies different sets of phase
shift parameters to each signal.  The beamformer chip therefore has
four inputs for four Vivaldi antennas, and two independent outputs.
The independent outputs is what makes it possible for EMBRACE to
have two independent fields~of~view, often referred to as
``RF~Beams'', which are named Beam-A and Beam-B.

The output of 3~beamformer chips is summed together on a ``hexboard''
and 6~hexboards make a tile. The EMBRACE array at \nancay\ has a
further analogue summing stage with 4~tiles making a tileset.  This
final stage is done on the Control and Down Conversion (CDC) card in
the shielded container which is connected to the tiles via 25~m long
coaxial cables.  

\newlength{\cellwidth}
\setlength{\cellwidth}{0.15\textwidth}
\begin{table*}[t]
\centering
\caption{\label{table:components}Components in the beamforming hierarchy.  The final component in the hierarchy of analogue beamforming is the CDC card.  There are two sets of 16 CDC cards, one for RF~Beam-A, and the other for RF~Beam-B.}
\begin{tabular}[h]{|r|c|c|c|}
  \hline \multicolumn{1}{|c|}{\bf component}
  & \multicolumn{1}{|p{\cellwidth}|}{\bf\centering number summed per group}
  & \multicolumn{1}{|p{\cellwidth}|}{\bf\centering summed by}
  & \multicolumn{1}{|p{\cellwidth}|}{\bf\centering total number per system}\\
  \hline & & &\\
  Vivaldi antenna & 4 & beamformer chip & 4608\\
  beamformer chip & 3 & hexboard & 1152\\
  hexboard & 6 & tile & 384\\
  tile & 4 & CDC card & 64\\
  CDC card & 16 & RSP boards & 16 (32)\\
  & & &\\
  \hline 
\end{tabular}
\end{table*}

\subsection{Control and down conversion}
\label{sec:CDC}
The CDC cards are responsible for three
important tasks \citep{bianchi_limelette, monari_limelette}:
1)~Frequency mixing the Radio Frequency (RF) for conversion to a
100~MHz bandwidth centred at 150~MHz; 2)~48~Volt Power distribution to
the tiles; 3)~Distribution of command and housekeeping data
communication to the tiles.  The RF, 48~Volt power supply, and
ethernet protocol monitoring and control communication are all
multiplexed on the coaxial cables connecting the tiles to the CDC
cards~\citep{berenz_limelette}.

The frequency down-conversion is done by heterodyne mixing in two
steps.  A first Local Oscillator (LO) is mixed with the RF signal of
interest which is between 500~MHz and 1500~MHz.  This LO is tuned
within the range of 1500~MHz to 2500~MHz such that the upper side band
frequency is precisely 3000~MHz.  The resulting signal is in turn
mixed with a second LO which is fixed at 2850~MHz resulting in the
Intermediate Frequency (IF) centred at 150~MHz.

There are 3~signal generators used to supply the LO signal to the
mixers in the CDC cards.  The LO at 2850~MHz is split by a
3~dB power splitter to supply the Beam-A and the Beam-B system.  This
signal is in turn divided by a cascade of 3~dB power splitters to be
distributed amongst the 16~CDC cards.
Similarly, the tunable LO is split amongst 16~CDC cards.  There is a
separate signal generator providing the tunable LO for the Beam-A and
the Beam-B electronics (16~CDC cards for each).

The fact that the LO is distributed amongst the CDC cards is a
potential source of an undesirable artefact in the final signal.  The
distributed LO signals could radiate directly into the CDC cards, or
possibly mixing products could have a component which is correlated
between all the outputs of the CDC card leading to a correlator offset
in the beamformed data of the whole array.  This is discussed further
in Sect.~\ref{sec:corroffset}.

\subsection{Digital processing}
\label{sec:digproc}
The output of the tilesets is fed into a LOFAR-type digital Receiver
Unit (RCU) and Remote Station Processing (RSP) system for digital
beamforming \citep{picard_limelette}.  The RSP performs the digital
beamforming of the entire array, producing pencil beams which are
usually called ``digital beams''.  These digital
beams can have pointings within the RF-beam produced by the individual
inputs to the RCU.  For \enancay, the inputs to each RCU is the signal
from a tileset with a beam width of approximately 8.5\degrees\ (see
Fig.~\ref{fig:tsbeam} and Sect.~\ref{sec:BeamPattern}).

Digital beamforming as described above is done in order to reduce the
volume of data produced by the system.  The alternative method of
producing the antenna cross correlations has the advantage of
providing a spatially fully sampled image of the entire field of view
every data dump (5.12\microsec).  This however produces an enormous
volume of data which cannot be managed by the acquisition system.  As
a result, the phased-array technique is used in the digital
beamforming to produce multiple beams on the sky, not necessarily
covering the entire field of view.

EMBRACE is a single polarization instrument but the LOFAR RSP system
has the capacity to produce two outputs per digital beam which
correspond to the two orthogonal linear polarizations in LOFAR.  These
are called the ``X'' and ``Y'' beams.  For EMBRACE, ``X'' and ``Y''
are two, possibly different, pointing directions within the field of
view (the RF~beam).  This means that for each digital beam, there are
two directions, and there can be any number of digital beams as long
as the total bandwidth remains within the limit of RSP processing
capacity (36~MHz for each RF~beam for \enancay).  The dual pointing
per digital beam is discussed further in Sect.~\ref{sec:M33}.

Fast data acquisition from the RSP boards is done by the backend
called the Advanced Radio Transient Event Monitor and Identification
System (ARTEMIS) developed at Oxford University
\citep{2015MNRAS.452.1254K, 2013IAUS..291..492S, 2012ASPC..461...33A}. ARTEMIS is a combined software/hardware
solution for both targeted observations and real-time searches for
millisecond radio transients which uses Graphical Processing Unit
(GPU) technology to remove interstellar dispersion and detect
millisecond radio bursts from astronomical sources in real-time.  The
pulsar observations with \enancay\ are targeted at known pulsars
applying known Dispersion Measure and timing parameters, and for this
reason the GPU are not used on the \enancay\ ARTEMIS backend.

The ARTEMIS hardware is also used for recording raw data packets from
the RSP boards containing the digitized beam formed wavefront data.  This
is used for the high spectral resolution observation of the extra
galactic source \object{M33} (see Sect.~\ref{sec:M33})

\subsection{Statistics data}
\label{sec:statsdata}
In addition to the high rate beamformed data produced by the RSP system,
there are also slower cadence data produced at a rate of once per second:
the crosslet statistics, the beamlet statistics, and the sub-band statistics.

The LOFAR RCU system digitizes and channelizes the 100~MHz wide RF bandpass
into 512 so-called sub-bands, each of 195.3125~kHz bandwidth. The cross
correlations of all tilesets are calculated once per second and are called
crosslet statistics. The default mode of operation for LOFAR is to calculate the
crosslet statistics for each sub-band in succession such that it takes 512 seconds to
cycle through the full RF band.

Another possibility is to request a given sub-band, and the crosslet
statistics are calculated each second for the same sub-band. This is
the mode of operation used most often at \enancay. The sub-band
statistics is simply the sub-band total powers of the 100~MHz bandpass for each tileset. The
beamlet statistics are the beamformed total powers for the full array
(16~inputs for \enancay) dumped at 1~second intervals, and are
identical to the fast data integrated over 1~second.

\subsection{EMBRACE monitoring and control software}
\label{sec:MAC}
The Monitoring and Control software for EMBRACE was developed at
\nancay.  An extensive Python package library on the SCU (Station
Control Unit) computer gives scripting functionality for users to
easily setup and run observations depending on various parameters such as, for example, the
type of the target, pointing, frequency selection, etc.  Integrated statistics data are acquired from the 
Local Control Unit (LCU) and saved into FITS files including header
information with essential meta data including pointing information,
timestamp, frequencies, etc.  Raw data (beamlets) are captured from
LCU Ethernet 1~Gbps outputs and saved into binary files
\citep{renaud_MAC}.

\section{Calibration}
\label{sec:cal}

\subsection{First stage phase calibration}
\label{sec:RFcal}
\begin{figure}[ht!]
 \centering
 \includegraphics[width=0.9\linewidth,clip]{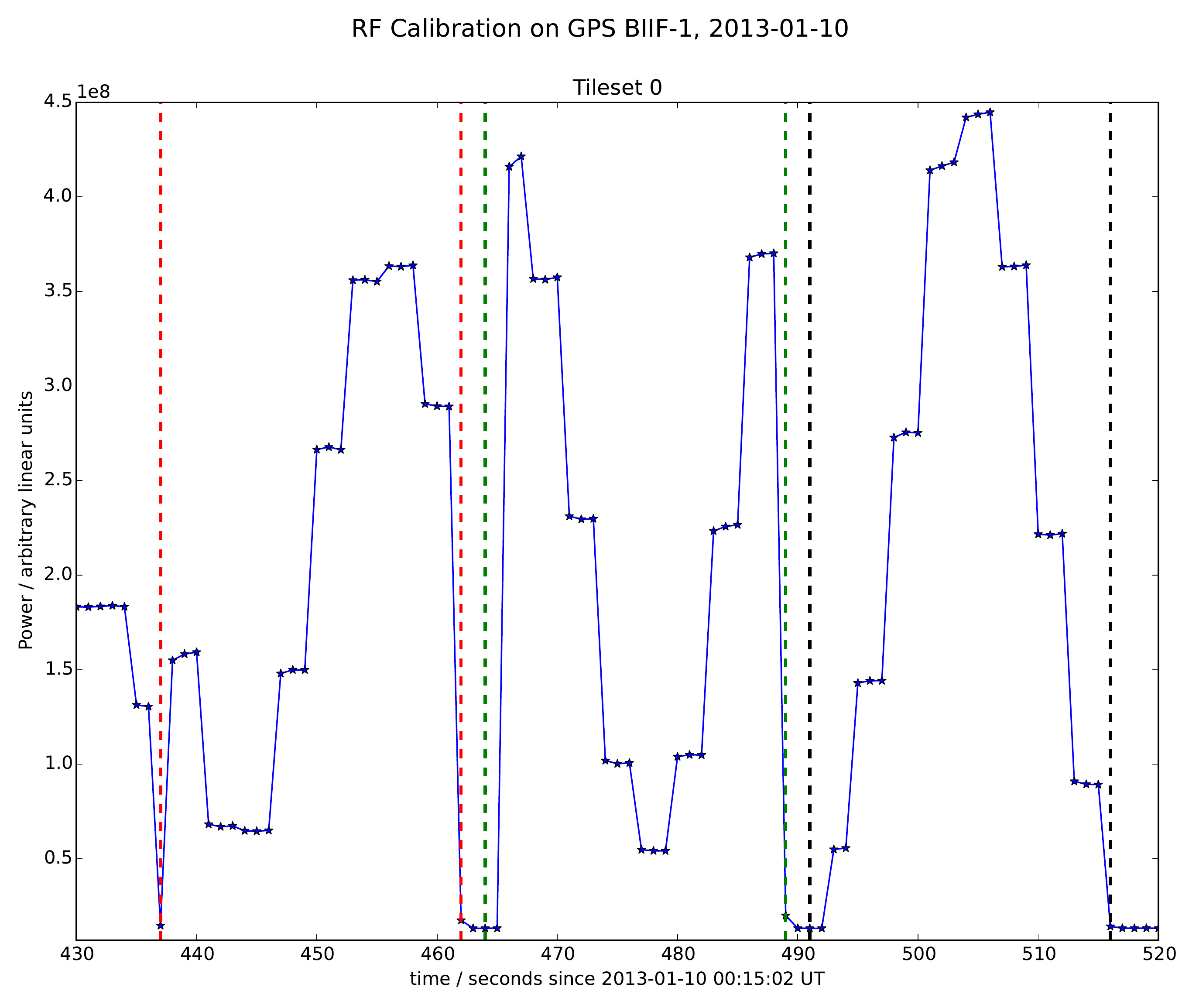}
  \caption{An example output of the RF calibration scheme for one tileset as described in the text.  The response of tile pairs for each of the 24~phase steps applied to the second tile in the pair is shown within the three groups of vertical dashed lines.  The response is clustered in groups of three corresponding to the constraint of the phase shift step size, however within each group of three there is some variation.}
  \label{fig:RFcal}
\end{figure}

\enancay\ has a hierarchy of four analogue beam forming stages. The
first three are done on the tile boards, while the last one is done on
the Control and Down Conversion (CDC) card in the shielded container.
The cables running from individual tiles to the CDC cards are 25~m in
length, and there are phase perturbations between the various
connectors and length of cable leading from each tile.  This is
calibrated out using an algorithm implemented in the Local Control
Unit\refereedelete{(see Fig.~\ref{fig:RFcal})}.

The phase correction between tiles in a tileset is measured by
successively maximizing the output power from pairs of tiles.
Initially, the required phase shift for each of the 72~antennas in the
tile is calculated from purely geometrical considerations (\ie\
antenna position and desired pointing direction).  This results in a
value between 0~and 360\degrees, different for each antenna.
Afterwards, an additional phase shift is added to all the antennas in
the tile in order to compensate the imperfections in the components
and cables.  Finally, the phase shift to be applied to each antenna is
quantized to 45\degrees\ steps.

This procedure is done in 24~phase steps, each time incrementing the
additional phase shift to be added to all the antennas in the tile by
15\degrees.  The step size of 15\degrees, followed by quantization to
45\degrees\ required by the beamformer chips, results in a slightly
improved configuration for the tile.  For example, if the phase shift
required by geometry for a given antenna is 7\degrees, and an
additional phase shift of 15\degrees\ is added, the result will be
22\degrees, which is quantized to 0\degrees.  A nearby neighbour might
require a phase shift of 8\degrees, and an additional shift of
15\degrees\ would result in a quantized step of 45\degrees.  In this
way, the quantization as the final step after application of a global
phase shift of 15\degrees\ allows for different configurations of
phase shifts amongst the 72~antennas.

The phase shift which gives the maximum result is then
converted to a value from 1~to~8 corresponding to the 8~phase steps
available in each beamformer chip.  This process is repeated for each
of three pairs of tiles in a tileset, with one tile acting as a
reference in each case (Tile~0).  At each iteration, two tiles are set
in ``no output'' mode while the other two make observations at each of
24 phase shift settings (see example in Fig.~\ref{fig:RFcal}).  The
calibration scheme is done in parallel for each of the 16~tilesets
which make up the \nancay\ EMBRACE station.

Since the analogue beam steering is based on signal phase shifts and
not true time delays, the corrections found are only strictly
valid for the centre frequency.  This does introduce an error in the
off-centre sub-bands but it is small since the bandpass is narrow
compared to the RF frequency, and it is mostly compensated in the
digital phase calibration which is the next step in the observation.

A strong source is necessary for this calibration procedure because
the source must be detectable by individual tiles, each with just over
1~m$^2$ of collecting area.  For EMBRACE, the Sun is often used as a
calibration source, as well as GPS satellites.  In particular, the
GPS~BIIF series of satellites which use the L5 carrier centred at 1176.45~MHz are good
calibration sources.  Only the sub-bands that receive significant power
will be usefully calibrated, but the commonly used GPS L5 carrier is
broad enough to calibrate at least 61 sub-bands ($\sim12$~MHz, one full
``lane'').

Once the necessary phase offsets for each tile have been measured,
they are stored in memory and used for the subsequent observation.
The phase offsets are also written to disk and can be used in future
observations without going through the calibration procedure described
here.  The use of calibration tables is the default mode of operation
and has been used successfully without updating the tables over
periods of many months (see for example the pulsar observations
described in Sect.~\ref{sec:psr}).

The measured phase offsets are different for different pairs of tiles,
and have values within the full range of 0~to~360\degrees which
corresponds to apparent path length errors of up to half a wavelength
($-\lambda/2<\Delta_{\mbox{path length}}<\lambda/2$).  For example, at 970~MHz, the path
length error for some pairs of tiles is as much as $\sim15$~cm.

\begin{figure}
\centering
\includegraphics[width=0.9\linewidth,clip]{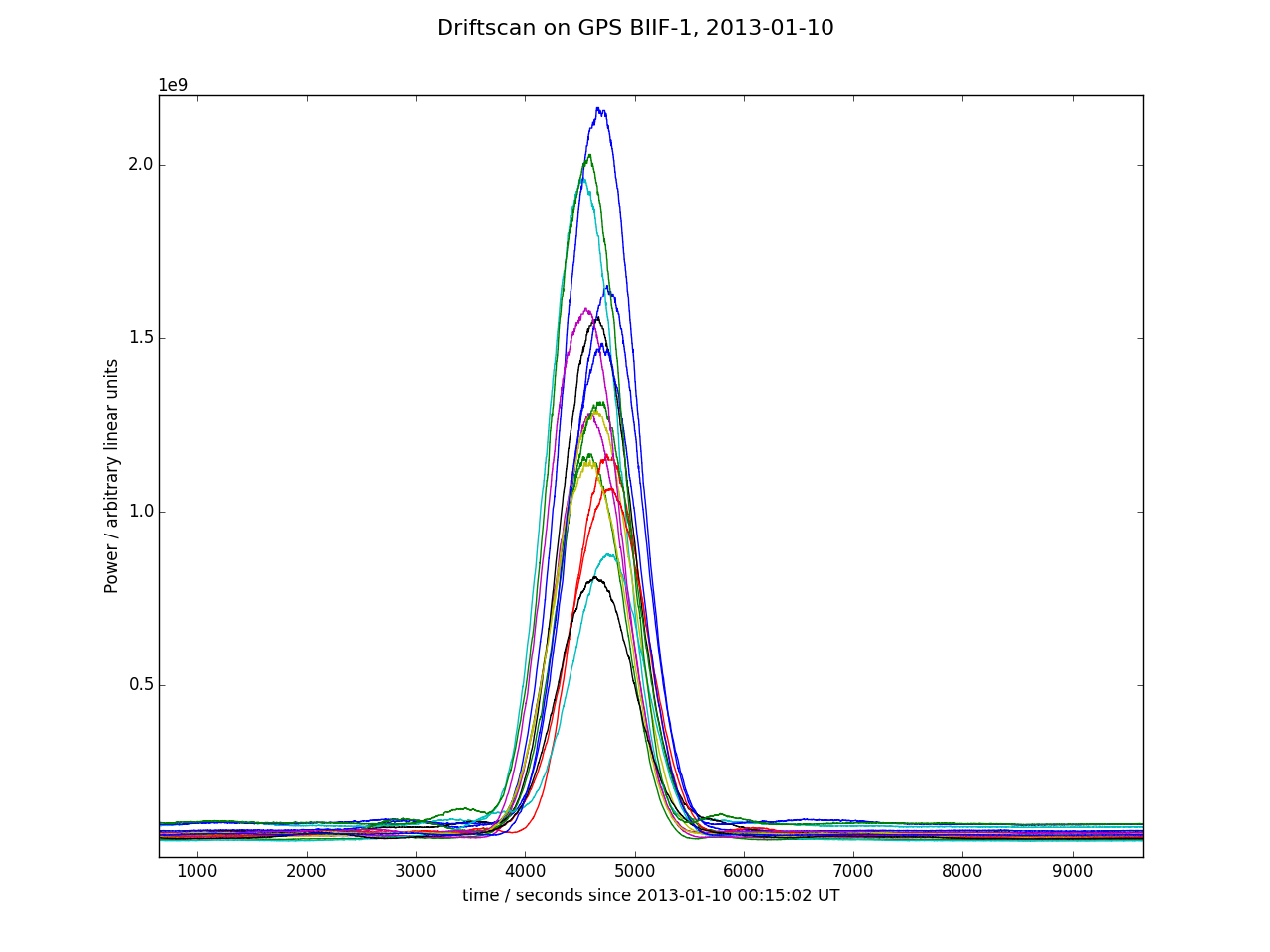}
\caption{Drift scan signals of the GPS BIIF-1 L5 carrier 2013-01-10 at
  80$^\circ$\ elevation in the north. Each line corresponds to a tile
  set and their half-widths, combined with the spacecraft's drift
  rate, can be converted to a HPBW beam size of $\sim8.5^\circ$ which is close
  to what is suggested by the standard formula
  1.2$\times$$\lambda$/$D$\ where $D$\ is the tileset side of 2.1~m.
  The drift peaks are aligned as a result of applying the necessary
  phase shifts to each tile.}
\label{fig:tsbeam}
\end{figure}

Figure~\ref{fig:tsbeam} shows a drift scan of satellite
GPS~BIIF-1 and each line is the output from a different tileset.  As
can be seen, each tileset is phased-up correctly, and they all show
the satellite peak at the expected time.  Gain equalization has not
been applied, but may be implemented in the future (see Sect.~\ref{sec:redcal}). The data here are
from the Sub-band Statistics.  For more details on this observation,
see Sect.~\ref{sec:BeamPattern}.

\subsection{Second stage phase calibration}
\label{sec:digcal}
\newcommand{\ewiseprod}{\;\odot\;}

The RSP digital electronics of EMBRACE described in
Sect.~\ref{sec:digproc} produce high speed beamformed data by
combining the complex waveform voltages of each tileset and multiplying by
a complex steering vector which gives the beamformed output in a desired
pointing direction.  The goal of the phase calibration is to determine
the modification to the complex weights required to compensate for any
instrumental offsets.

\modified{
In order to describe the procedure for determining the phase
calibrated pointing, matrix operations are used including the matrix
dot product (denoted by the central dot) and element-wise
multiplication of matrix elements, also called the Hadamard product, denoted by $\ewiseprod$.  Complex
conjugate of a matrix is denoted by a superscript asterisk and the
transpose with a superscript $T$.
}

The RSP system produces the high cadence beamformed \modified{voltages by forming
the dot product of a complex steering row-vector $w$ with the complex
voltage column-vector $T$ received per tileset:}
\begin{equation}
V=w \cdot T
\label{eq:digbeamforming}
\end{equation}

The steering vector depends on the array geometry given by the tileset
positions, and on the desired pointing in the sky given by the
direction cosines \citep[see for example][]{interfero_bible}.  \modified{The
steering vector must be modified further to take instrumental
offsets into account.}

The calibration procedure makes use of the fact that beamforming in a
given direction can be done equivalently by combining the steering
vector with the cross-correlation matrix~$X$ which is the correlations
between tilesets, and this produces the beamformed power $P$,
\begin{equation}
  P=w\cdot X \cdot w^*
  \label{eq:skymap}
\end{equation}
\modified{and we maximize $P$ while pointing at a calibration point source.}

The cross correlation between tilesets is calculated each second as
described above in Sect.~\ref{sec:statsdata}.  This is not fast
enough to produce the high speed beamformed data at 5.12\microsec\
intervals, but the cross correlation data can be used to estimate
phase errors between each tileset, per sub-band.  Each second, the RSP
returns a matrix with the cross correlations between tilesets at a
given frequency sub-band.

If the tilesets are pointing at a strong point source, then it is
possible to calculate an additional phase correction matrix which,
when multiplied \modified{element wise} by the measured correlation matrix $X$, results in the
expected point source at the centre of the field of view.  For an
ideal instrument, this phase correction matrix would be simply the
identity matrix.  In a real system, there are various
disturbances to the phase along the signal chain which varies from
tileset to tileset (cables, filters, amplifiers, mixers, analogue-to-digital
converters, etc).  These phase differences must be measured using a
strong source in the sky, and calibrated for future observations.

We look for a phase correction matrix $W_{cal}$ such that \modified{the element-wise product}
\begin{equation}
  X_{cal}=X\modifiedmath{\ewiseprod} W_{cal}
  \label{eq:calvisibilities}
\end{equation}
\modified{results in the matrix $X_{cal}$ which}
is the corrected correlation matrix fixed according to
a theoretical model of centred point source.  \modified{This means that the $X_{cal}$ elements are all real.} $W_{cal}$ may be further decomposed into a steering
matrix $W_{\modified{geom}}$ giving the phase required to steer the beam in a given
direction given the array geometry, and a correction matrix $W_{cor}$
including the corrections of all other sources of phases errors. We
then have
\begin{equation}
  X_{cal}=X\modifiedmath{\ewiseprod} W_{\modified{geom}}\modifiedmath{\ewiseprod} W_{cor}.
  \label{eq:weightmatrix2}
\end{equation}
The phase relations between elements at a given $i,j$ is thus
\begin{equation}
  arg(X_{cal}^{i,j})=0=arg(X^{i,j})+arg(W_{\modified{geom}}^{i,j})+arg(W_{cor}^{i,j}).
\end{equation}

$W_{cor}$ can then be computed by a simple element wise division:
\begin{equation}
  W_{cor}=X_{cal}/(X\modifiedmath{\ewiseprod}W_{geom})
\end{equation}
which corresponds to the phase relation for given $i,j$ indices:
\begin{equation}
  arg(W_{cor}^{i,j})=\;-arg(X^{i,j})-arg(W_{geom}^{i,j})
\end{equation}

A complete phase correction matrix $W_{cal}$ can then be created with magnitude unity and
with the phases required to correct the measured cross correlation
matrix, including the steering step and the phase correction to be applied:
\begin{equation}
  W_{cal}=W_{\modified{geom}}\modifiedmath{\ewiseprod} W_{cor}.
  \label{eq:phasecorrectormatrix}
\end{equation}

In order to use the calibrated steering matrix for the high cadence
beamforming, it is necessary to convert it back to a steering vector.
\modified{
This is done according to the relation
\begin{equation}
  W_{cal}=w_{cal}^T \cdot w_{cal}
  \label{eq:steeringtmatrix}
\end{equation}
and $w_{cal}$ can be} extracted using the method of
Singular Value Decomposition \citep[see for example][]{noble_svd}.  It
is this calibrated steering vector which is used by the RSP system to
calculate the EMBRACE beamformed data at 5.12\microsec\ intervals using
Eq.~\ref{eq:digbeamforming}.

As with the calibration of the tilesets described above in
Sect.~\ref{sec:RFcal}, these calibration parameters are saved to
disk and can be used for future observations without the necessity of
making a calibration measurement before each observation run.  The
calibration table measured once has been valid for many months, as
discussed further in Sect.~\ref{sec:psr}.

\subsection{Redundancy gain calibration}
\label{sec:redcal}
A further digital calibration method based on the redundancy
method \citep{red1} is used in post processing.  This method is
particularly well suited to EMBRACE given the inherent level of
baseline redundancy.  \enancay\ is made of 16~tilesets placed on a
regular $4\times4$~grid and therefore one point in the UV space is
measured by several different baselines sharing a common length and
orientation.

The problem becomes one of solving an ill-conditioned set of equations
in which there are more equations than there are parameters
to be determined.
\begin{equation}
X_{ij}^{obs} = X_{ij}^{true}w_{i}w_{j}^{*} + e_{ij} 
\label{eq:red}
\end{equation}
where $X_{ij}^{obs}$ and $X_{ij}^{true}$ are respectively the observed
and true cross correlations corresponding to baseline $ij$; $w_i$ is
the complex gain of tileset i; and $e_{ij}$ is the noise affecting
baseline $ij$.  One solves this system for all $X_{ij}^{true}$ and
$w_i$, with baseline redundancy giving a redundant structure to
$X_{ij}^{true}$ (several baseline outputs modeled by the same true
cross correlation).  As shown in e.g. \citet{red1} and \citet{red2},
there exist several ways of implementing a redundancy calibration
method which solves Eq.~\ref{eq:red}. We implemented a
non-linear method based on a classical steepest gradient algorithm
\citep[see~e.g.][]{red3}.

In addition to phase calibration, the redundancy method allows for an
estimation and correction of gain amplitude disparity between
tilesets, which is the main reason for its implementation.  This
equalization of tileset gains allows for optimization of beam shape
features like symmetry, resolution and stability over time (see
Sect.~\ref{sec:beamsim}).  It can also be used to clean the data
from unwanted instrumental effects by comparison with an ideal square
aperture model.  Note however that this method is not intended to
optimize the \SNR\ which may require a different weighting of tilesets
based on individual \SNR\ ratios and side lobe levels.

The redundancy calibration is robust since it relies
only on the similarity of the individual elements. The calibration quality depends only on the \SNR.  The
point-like or extended nature of the source does not affect the
result.  It nevertheless requires the position of the source as an
input for an absolute phase calibration, since the resolution of
equation~\ref{eq:red} is invariant with respect to a global shift of
the sky (absolute calibration of gain amplitude will not be dealt with
here). This can be done by adding dedicated constraints to the solver,
for example with the implementation of a phase correction gradient minimization across
the array, or the inclusion of a model of the observed source.
\citep{red1,red2}.

The only limitations are thus expected at low signal over noise, or
when individual elements have significantly different behaviour from
each other.  The latter cause may come from intrinsic inhomogeneities
in the numerous system electronic components on the array itself, the
mass production employing industrial techniques with good
reproducibility should nevertheless strongly mitigate this risk. It
may also arise from environmental effects, e.g. temperature
inhomogeneities in the electronic components over the array. The
self-heating of dissipating electronics seems however to ensure a
rather stable and homogeneous temperature over the electronic boards
that form the array.  Finally mutual coupling between antennas could
also be a source of disparity among the elements. Nothing so far indicates
that such coupling is of importance on \enancay, but the
dense aspect of the array calls for caution on that point. An
application of redundancy calibration on a relatively dense array can
be found in \citet{red2} where data from a LOFAR HBA station are
analyzed near the critical wavelength $\lambda_{c}$ defined by the
minimum distance $d_{c}$ between individual antennas. While those HBA
arrays have been designed as "semi sparse" across their entire
frequency band ($\frac{\lambda}{2}<d_{c}<2\lambda$), EMBRACE has
access to both semi-sparse regime ($F>1200~{\rm MHz}$) and truely
dense regime ($F<1200~{\rm MHz}$). The possible mutual coupling
effects will be investigated with further development of the system model
(see Sect.~\ref{sec:beamsim}), if needed a refinement of
the model in Eq.~\ref{eq:red} could be implemented to take mutual coupling
into account.

\refereedelete{
The criteria on individual element similarity and \SNR\ may
be a concern. The similarity of individual elements may be perturbed due to
intrinsic inhomogeneities of the system components or environmental
effects. Influences of the environment are
\textit{a priori} unknown, and may be classified in two major groups:
Basic environmental influences like temperature that may be coupled to
the array operating conditions, and mutual coupling between internal
elements.  

At frequencies below 1200~MHz, EMBRACE is a densely packed array of
antennas spaced by less than half the wavelength.  Mutual coupling
between antenna elements is significant compared to a sparse
array. This coupling is known to affect the similarity of the element response.  Mutual coupling might result in excess noise,
or higher side-lobe levels. It may also restrict the validity of the
redundancy solution, necessitating a refinement of the model in
equation \ref{eq:red} if mutual coupling is too high. An application
of redundancy calibration on a relatively dense array can be found
in \citet{red2} where data from a LOFAR HBA station are analyzed near
the critical wavelength $\lambda_{c}$ defined by the minimum distance
$d_{c}$ between individual antennas. While these HBA stations have
been designed as ``semi-sparse arrays'' across their entire frequency
band ($\frac{\lambda}{2}<d_{c}<2\lambda$ for any $\lambda$ in the
band~\citep{densparse}, EMBRACE has access to both semi-sparse regime
($F>1200~{\rm MHz}$) and truely dense regime ($F<1200~{\rm MHz}$).
}

\section{Correlator offset}
\label{sec:corroffset}
\label{sec:embdat}

\subsection{Description of the problem}
\begin{figure}
\centering
\includegraphics[width=0.95\linewidth,clip]{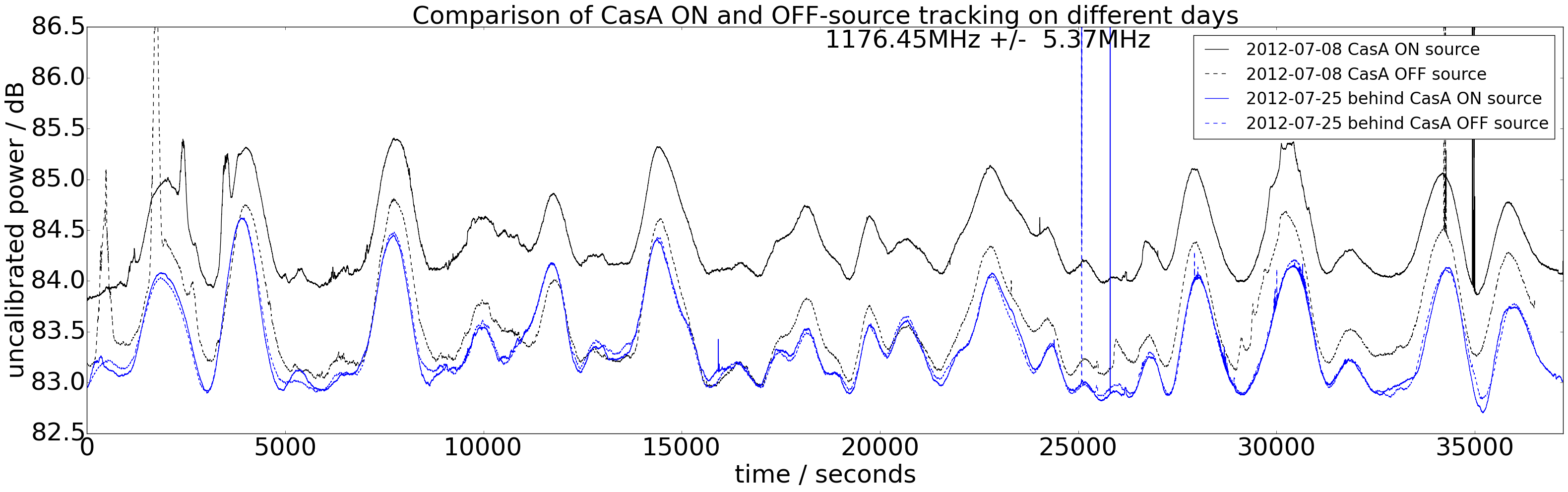}
\caption{Power as a function of time while tracking \object{Cassiopeia~A}, and tracking an
empty sky position with the same declination. Note that the solid black line is
the only on-source signal. Each tracking produces two curves due to the
dual digital beams used as illustrated in Fig.~\ref{fig:obsstrat}.}
\label{fig:casatrack}
\end{figure}

Long duration tracking observations of weaker sources revealed an
apparent power variation as a function of pointing direction that was
highly repeatable between observations. After some analysis of this
problem it was realized that the cross-correlation matrix data product contained
time-invariant complex offsets that are much larger than white noise
component when observing empty sky. As it turns out the `direction dependence'
is only an artefact of applying a varying set of complex weights (while
steering the beam to track a source) to a fixed non-changing pattern of
complex numbers. That description of the problem was observationally
confirmed by noting that the wobbly trace was present and exactly the same
whether we tracked a real source or just empty sky, as long as we followed
the same trajectory in the local sky. For strong sources,
such as the Sun or satellite carriers, this additive contribution is
completely negligible and does not become an issue, but by pure coincidence
the amplitude of the variation is comparable to the continuum power of
\object{Cassiopeia~A} and \object{Cygnus~A} (the two strongest extra-solar radio sources at
low frequencies).
This can be seen clearly in Fig.~\ref{fig:casatrack}.  It shows power versus
time from a \object{Cyg~A} tracking, and also tracking an empty position with the same
declination but different RA. The average power difference in these two signals
corresponds to the \object{Cyg~A} continuum level (the real astronomical signal) but
as is also evident, the magnitude of the \emph{variation} in the signals is
about of the same strength as the average power difference.

\subsection{Strategies for correcting the correlator offset}
\begin{figure}
\centering
\includegraphics[width=0.7\linewidth]{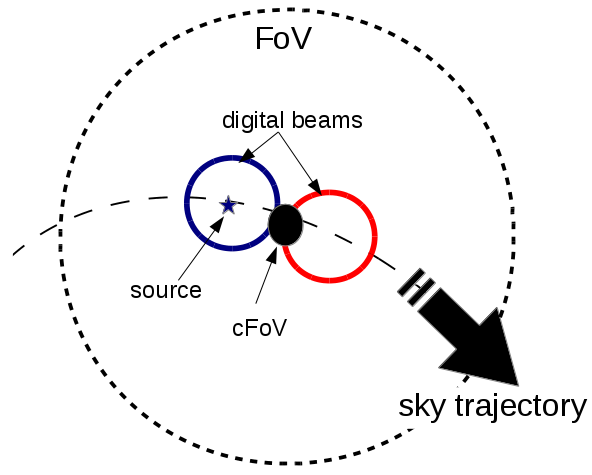}
\caption{Source tracking strategy.  Two digital beams are placed symmetrically within the FoV
(tileset beam) following the same sky trajectory but with a time delay.
One of them is centred on the source, the other is used as an off-position.}
\label{fig:obsstrat}
\end{figure}

An obvious quick method to get rid of the variation is to subtract the (solid)
blue and red signals shown if Fig.~\ref{fig:casatrack} and in order to have that
possibility, we routinely employ the scheme outlined in Fig.~\ref{fig:obsstrat}.
The LOFAR-type backend is always outputting two digital beams (see Sect.~\ref{sec:digproc})
and we use one of them to track the source, and the other to observe an off
position that follows the same trajectory on the sky.
Both beams are placed symmetrically within the tileset FoV so that the gains
within the two beams are the same.

\begin{figure}
\centering
\includegraphics[width=0.4\linewidth]{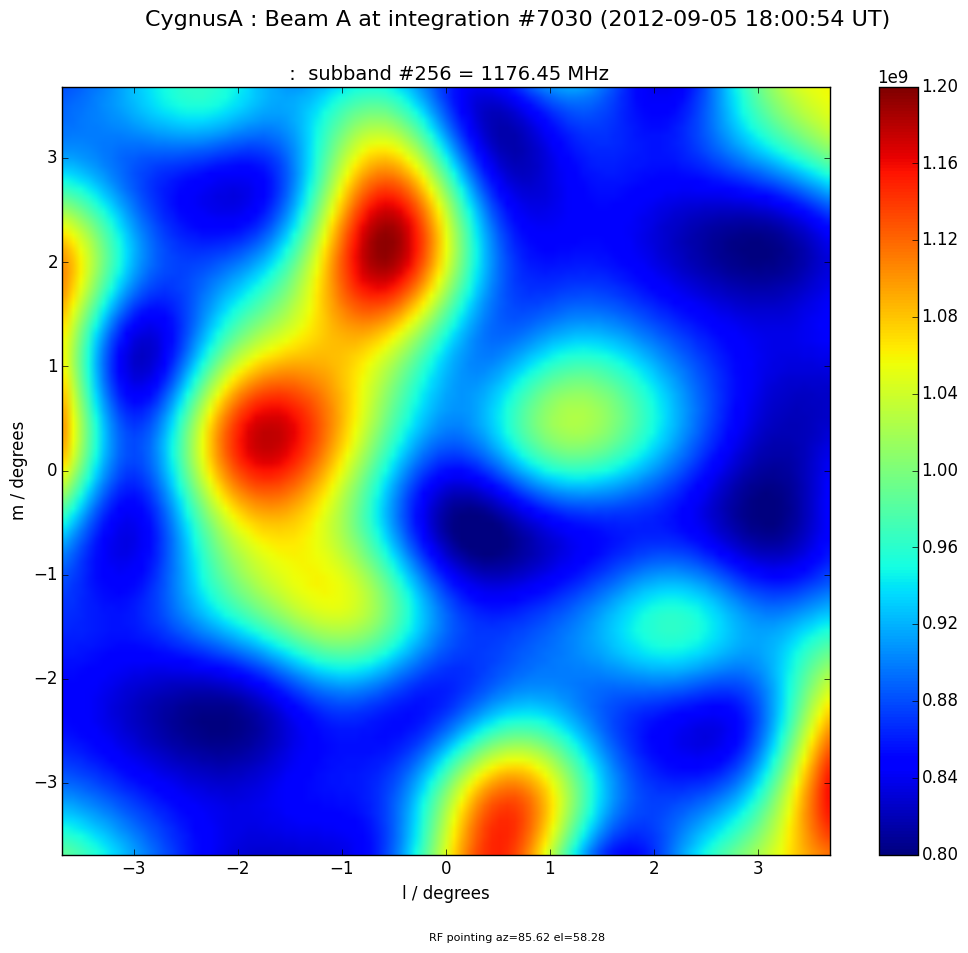}
\includegraphics[width=0.4\linewidth]{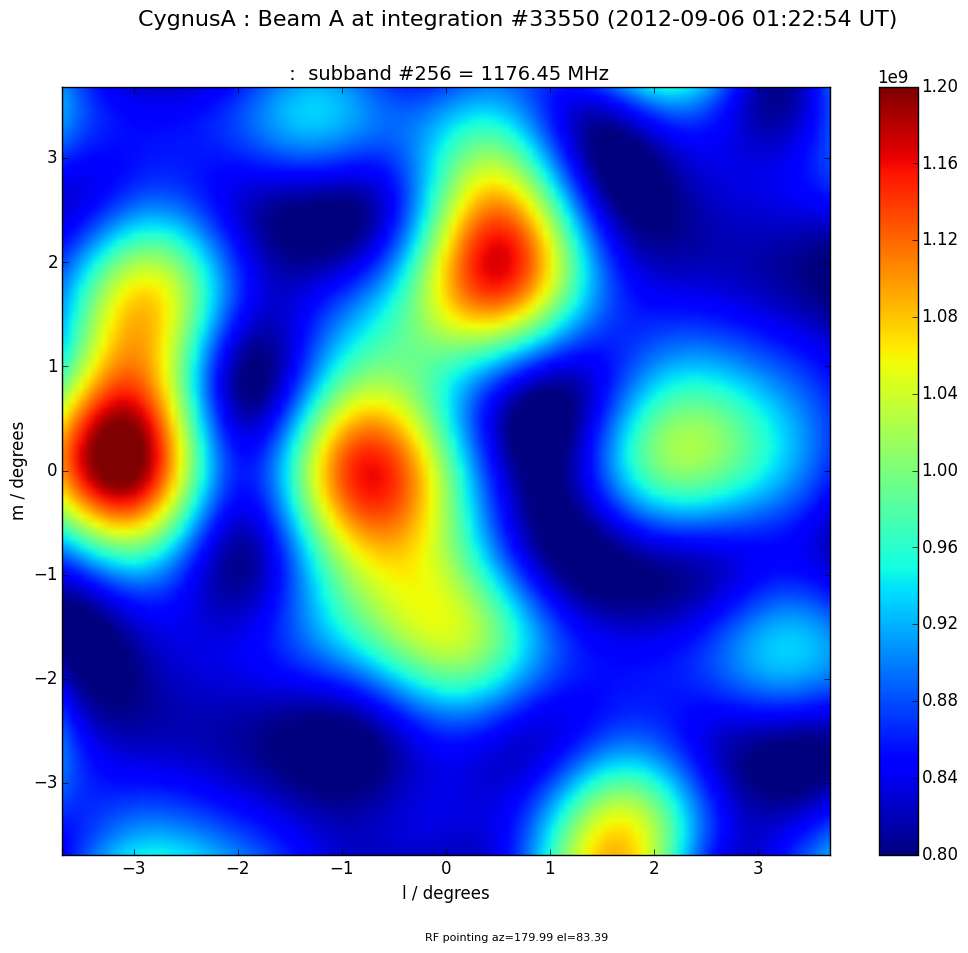}\\
\includegraphics[width=0.4\linewidth]{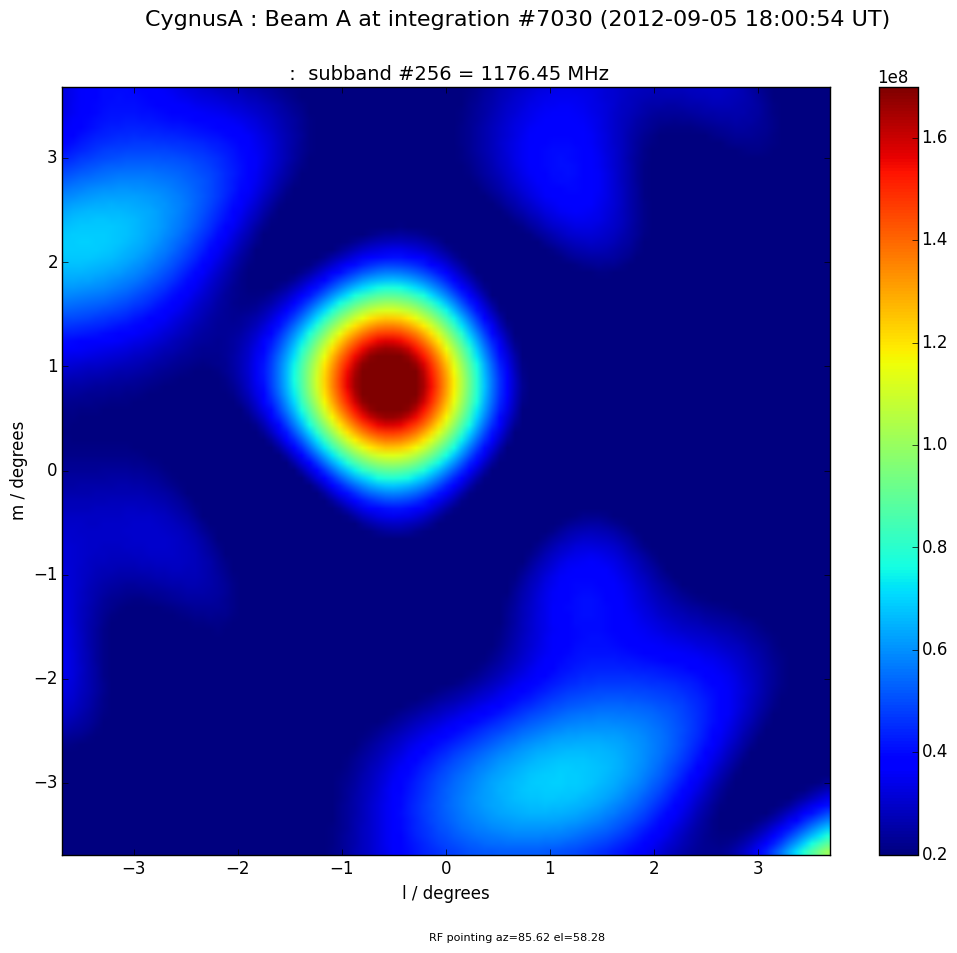}
\includegraphics[width=0.4\linewidth]{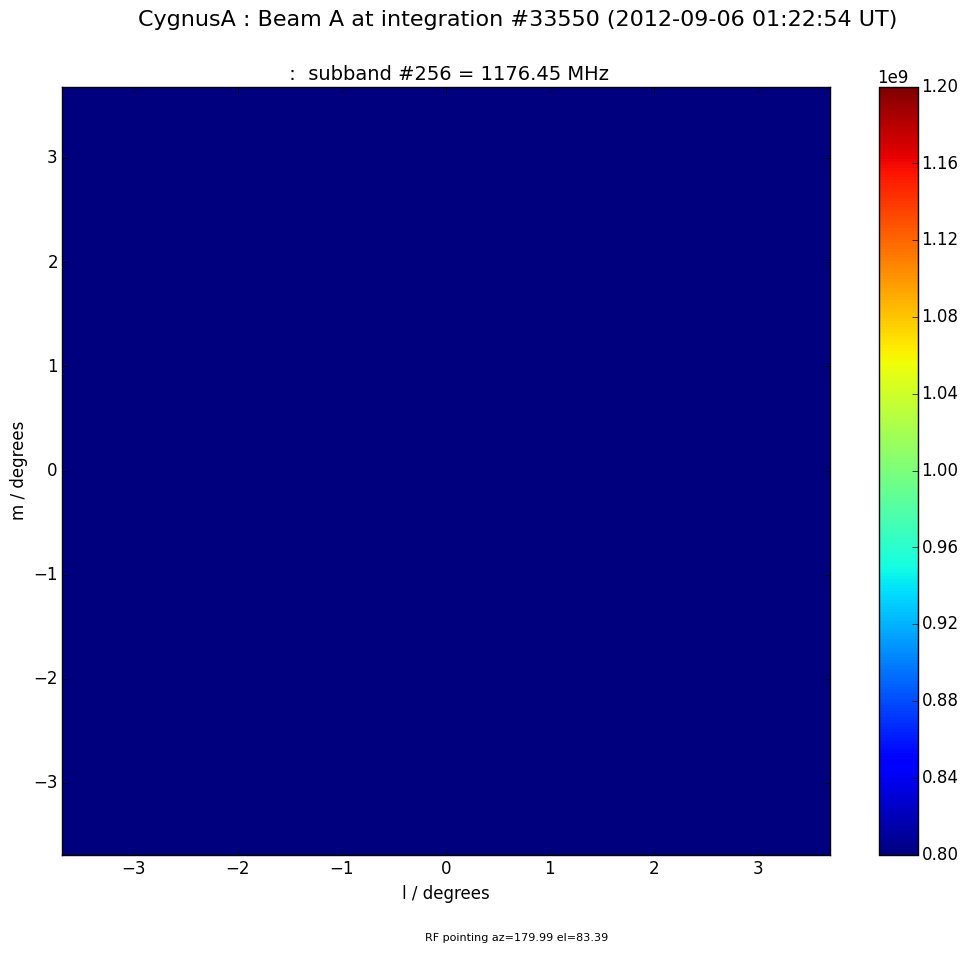}
\caption{Correlator offset seen in sky images.  The correlator offset problem discussed in the text is also manifested
by blob-like features when imaging the tileset FoV. Here the left and right
upper panels show FoV images of \object{Cygnus~A}, and towards an empty sky position,
based on uncorrected cross correlations. The bottom panels show
the same data but with images created from cross correlations corrected by
subtracting a reference matrix of complex constants (the same reference was
used for both images).
In the left bottom panel, \object{Cygnus~A} is now clearly seen and at its expected
position (offset from centre due to the scheme outlined in
Fig.~\ref{fig:obsstrat}).}
\label{fig:imfix}
\end{figure}

Such a measure is more similar to mitigating the symptom rather than
curing the cause of the problem. Cross correlations are easily corrected by
subtracting (per sub-band) a reference cross correlation slice measured towards
empty sky and this can be done in post-processing. We subtract all slices
in the dataset in question with the same reference slice that may
have been measured weeks or months earlier while tracking a different part of
the sky. With corrected cross correlations we can then create our own digital beams
in any direction within the FoV by applying the required complex weights.
This method cures the problem to a large extent as is demonstrated in
Fig.~\ref{fig:imfix}.

But there are severe limitations to this approach: the maximum cadence
for cross correlations is one matrix per second and then only for one
and the same sub-band. For $N$\ sub-bands, the cadence is decreased further to one
matrix every $N$\ seconds for each sub-band.

\label{sec:correctFastData}
The high cadence data stream is unfortunately not straight forward to
cure.  It consists of beamformed complex voltages every 5.12\microsec\
which is the vector product of a row weight vector and a column
tileset voltage vector (see Sect.~\ref{sec:digcal}
Eq.~\ref{eq:digbeamforming}). The complex offsets seen in the cross
correlations (conjugate products of tileset pair voltages) must in
turn be caused by complex offsets in the individual tileset signals
themselves. We then have
\begin{equation}
V_{\rm sb}=\mathbf{w}\mathbf{T}=\mathbf{w}(\mathbf{S}+\mathbf{E})
\label{eq:correctFastData}
\end{equation}
where $V$\ is the beamformed voltage for a given sub-band
$\mathit{sb}$, $\mathbf{w}$\ is the steering vector, $\mathbf{T}$\ are
the complex voltages received per tileset, $\mathbf{S}$\ is the true
uncorrupted sky signal from each tileset, and $\mathbf{E}$\ is an
error vector with complex constants.  It is not possible to correct
the complex voltages in post processing because they are the product
of beamforming which is done in real time.  We do not have access to
the individual tileset voltages for post processing.  It may be
possible to determine the error vector $\mathbf{E}$\ by using the
1-second cadence cross correlation data, but in order to use this, the
backend must calculate a subtraction before applying the complex
steering vector, and this is not possible within the current
processing power of the LOFAR-RSP system.

One can note that having a constant amplitude and phase value added to
the tileset signal for a given sub-band, that is changing in between
tilesets, is equivalent to injecting a tone into the system at the
particular frequency that is picked up differently per tileset
(modified amplitude and phase).  This ``tone'' -- that appears to be
always present as soon as the instrument is turned on, and which
is \emph{independent of the current analogue direction (beamformer
chip settings)} -- might be described as a waveform since the values
for neighbouring sub-bands follow one another logically. \refereedelete{This is
illustrated in Fig.~\ref{fig:polvis} which shows the complex
cross correlations in a polar plot for all sub-bands for one single baseline
while observing empty sky.}  It may also be of great importance to note
that LOFAR data, which are produced by a virtually identical backend
and deliver the same data products, do not at all suffer from this
effect. This suggests an origin in the analogue side of the instrument
which is of radically different design compared to LOFAR, one example
being that EMBRACE has an additional IF stage and subsequent
downmixing is using a common LO source for all CDC cards (see
Sect.~\ref{sec:CDC}).

\refereedelete{
\begin{figure}
\centering
\includegraphics[width=0.9\linewidth]{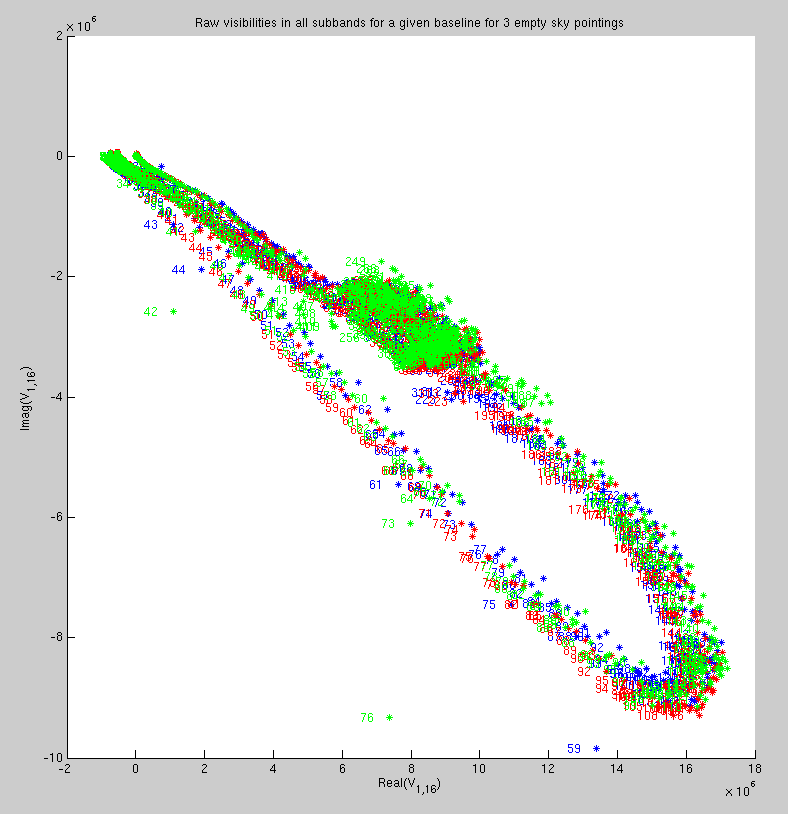}
\caption{Cross correlations for tilesets 1 \& 16, for all sub-bands at three
different times while tracking empty sky, plotted in the complex plane.
The numbers are the sub-band indices stretching from 1~to~512 across the 100~MHz
bandpass.}
\label{fig:polvis}
\end{figure}
}

\subsection{Investigation of the correlator offset}
\label{sec:corroffsetdetails}
We have previously noted that the fixed error pattern in the baseline 
correlations is stable enough to allow subtracting it out using a
template correlation matrix observed weeks or months earlier.
Furthermore, the template can be observed when the frontends are internally
phased up to an arbitrary direction on the sky.
The correction usually improves the beamformed continuum
time stability down to a level where the remaining variation is comparable
to what is seen while staring at a fixed point.

During an observation, an artificial ondulation in the beamformed power
trace is created by applying varying complex steering gains to static tileset
signals that are erroneuously correlated (see Fig.~\ref{fig:casatrack}).
In fact, there is no true temporal instability as one is first led to believe. 
Whereas the backend does not allow access to individual tileset signals in
the digital domain, we continue to employ the 1-second integration
correlation matrices to analyse the problem. Primarily we aim to understand
the origin and behaviour of the problem that we refer to as
``correlator offsets''.

\begin{figure}
\centering
\includegraphics[width=0.95\linewidth,clip]{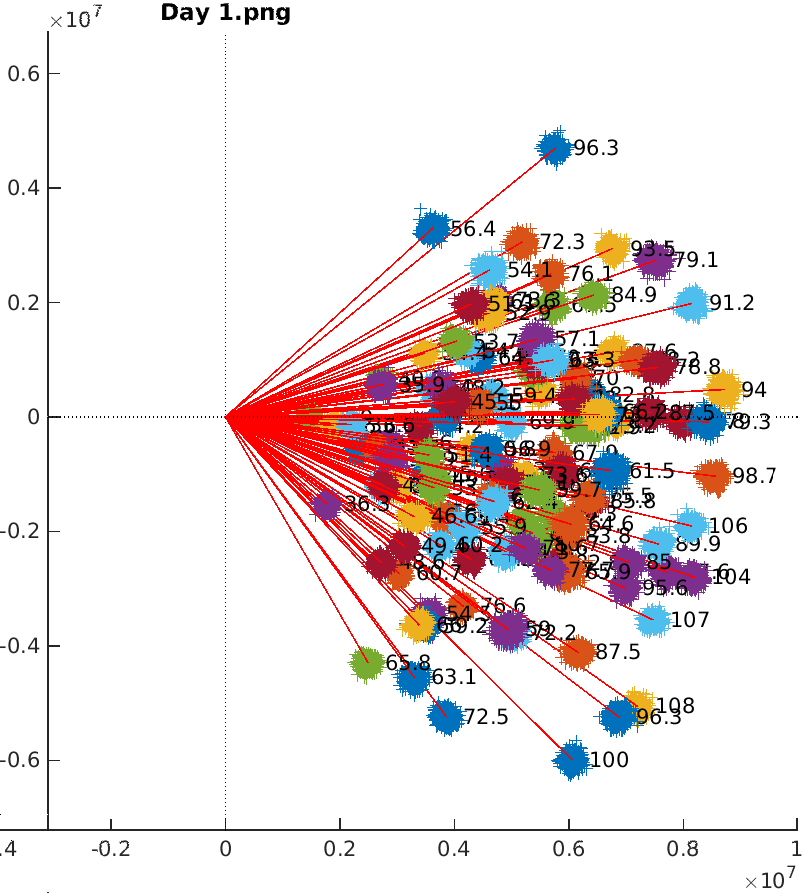}
\caption{A plot of the complex plane showing the correlator offsets while
observing empty sky.  The offset is much larger than the thermal
scatter of the points during 10~minute observations.  The red lines are simply to mark the separation from the plot origin.}
\label{fig:corroffset}
\end{figure}

In order to quantify the variability of the correlator offsets, we
have selected one sub-band in a 3-day data set where correlations for
all baseline pairs were sampled every second. The frontends were fixed
to the same analogue beam steering through the observation. The
observation was configured to let Cas~A drift through the beam once
per sidereal day thus regularly verifying that the system was
operating normally.  We chose portions of ten minutes every day,
separated by exactly 24 hours, well away from the Cas~A drift signal
and obvious transients, to study the daily fluctuations of correlator
offsets and compare them to the thermal noise as measured during each
600~second sequence (Fig.~\ref{fig:corroffset}).  We find that the
thermal 1-sigma scatter is 20--100 times weaker than the magnitude of
the complex offsets. The large range is due to the varying offset
magnitudes; the thermal scatter is rather constant over all baseline
pairs.  The magnitudes of the average daily change of the offsets were
found to be comparable to the thermal scatter, meaning that the
thermal scatter is 15--100~times smaller than the offset magnitude.

To study the long-term stability of the offsets, we make use of the
quasi-continuous pulsar tracking observations at 970~MHz that have
been conducted with \enancay\ since 2013. The pulsar signal is
effectively negligible in the baseline correlations and we pick one
correlation matrix per observation from $\sim400$~observations spread out
over 20~months.  Two different sub-bands were used and we first discuss
the batch which used the same sub-band continously for seven months.
During this time we find that the complex correlator offsets varied
somewhat -- up to 3-4 times the thermal scatter -- mostly in the
radial direction, indicating a fluctuation in input powers but not a
change in relative phases between tileset signals.  In
occasional observations all correlations with one particular tileset
were rotated exactly 90$^\circ$ indicating a corresponding jump in
phase for this tileset. At other observations exhibiting this effect,
it may have been another tileset that was affected.  Lastly, after
$\sim40$~days into the sequence (186 days in total) the LO powers were
permanently modified as part of a test and here we see a jump in the
positions of the offsets, again mostly in the radial direction.

\begin{figure}
\centering
\includegraphics[width=0.95\linewidth,clip]{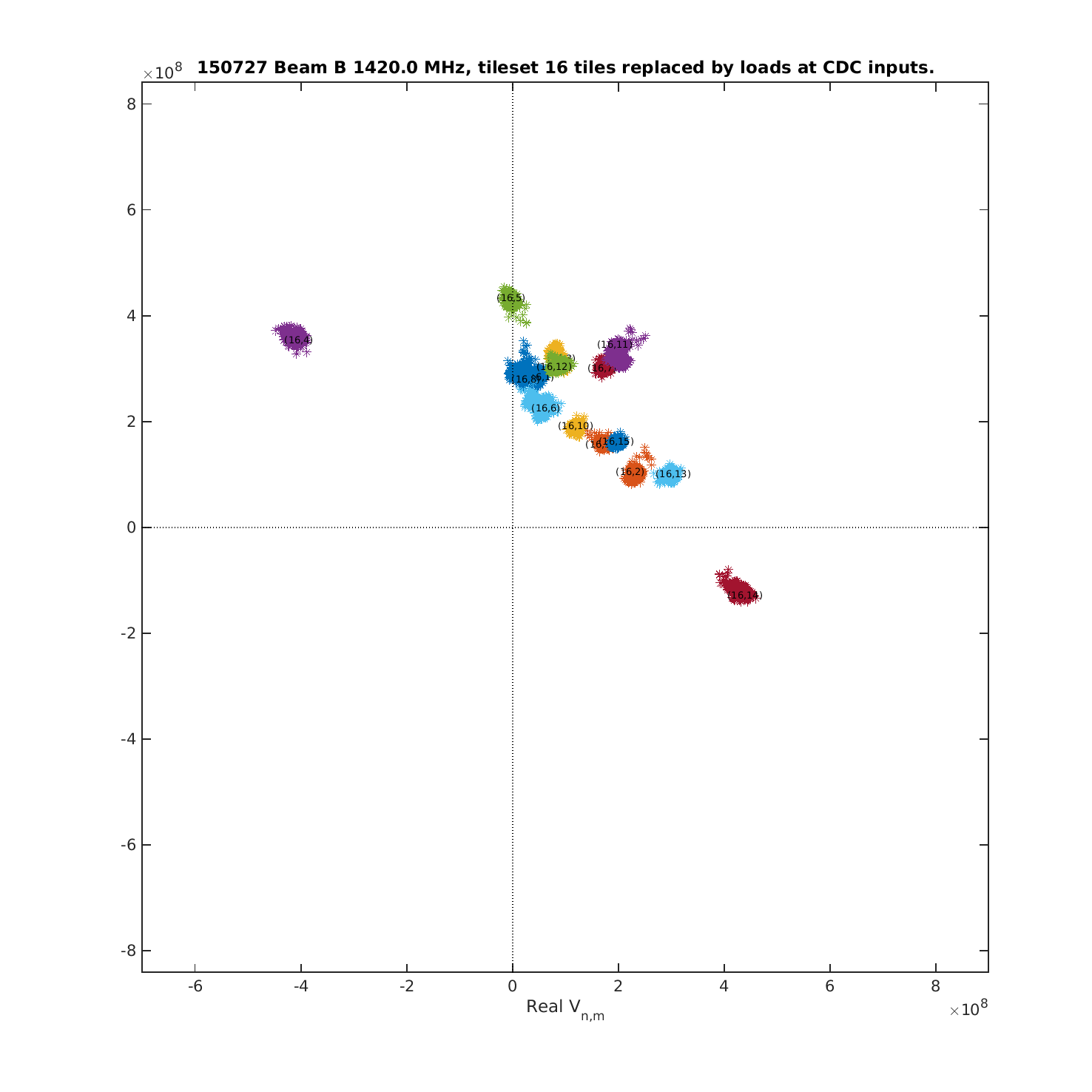}
\caption{Correlator offset without frontend.  When one tileset is replaced by a 50~Ohm load, the correlator offset remains, showing that the correlator offset is not due to the frontend system (antennas, amplifiers, beamformer chips, etc).}
\label{fig:loadcorroffset}
\end{figure}

In order to definitively confirm that the erroneous correlated signals
are not in any way related to the RF signals from the tiles
themselves, we disconnected the tiles for one tileset and instead
replaced the cables with termination loads directly at the inputs of
the associated CDC card (see Fig.~\ref{fig:loadcorroffset}) It turns out that this
did not affect the correlator offsets significantly, all correlations
with that tileset remained in the same place in the complex plane even
though the autocorrelation power level for that tileset clearly
changed (as it should since no effort was made to perfectly simulate
the normal tile output power levels).

On the other hand, turning off the LO amplifiers for all LO signals led
to the backend outputting correlations that were two magnitudes smaller,
and now spread symmetrically around the origin. Thus we conclude that
the offsets are \emph{not} produced by a malfunctioning backend, but
rather with the LOs or their mixing scheme or the CDC card electronics.

Comparisons were made with the second RF beam, ``Beam~B''.
It is configured the same way as Beam~A except that 20dB attenuators,
normally located just before the RCU inputs, were removed.
We see that fundamentally the problem exists in Beam~B as well, but here
the scatter in angles around the real axis is larger (ca +-90$^\circ$),
and crucially the magnitudes are almost exactly 100 times stronger.
The latter should provide another piece of evidence that the undesired
correlated signals are created before the RCUs (and the location of the
attenuators). The Beam~B offset magnitudes, when normalized by the Cas~A
signal strength, are still about two times larger than for Beam~A but this
is probably unrelated to the removal of the attenuators.

\begin{figure}
\centering
\includegraphics[width=0.95\linewidth,clip]{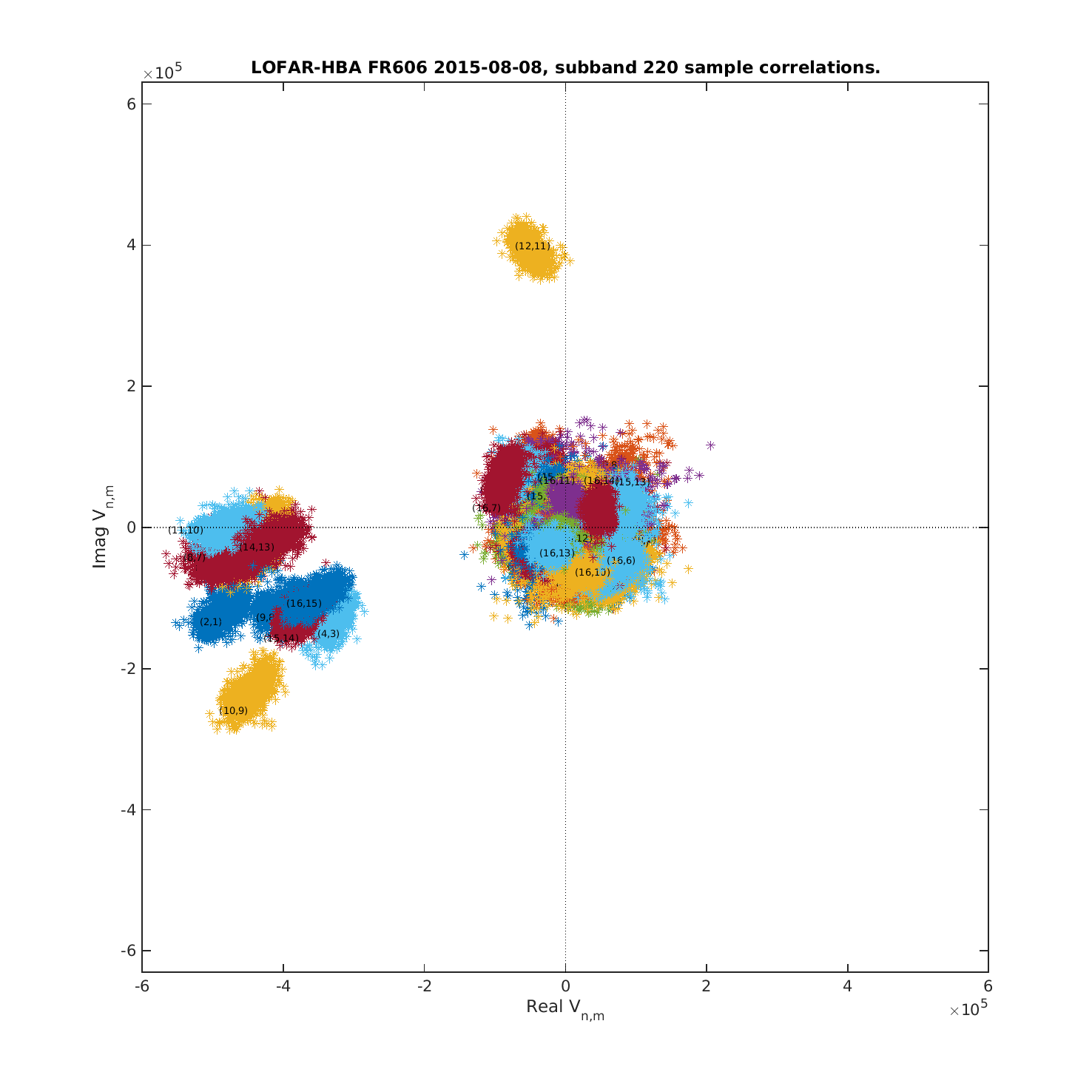}
\caption{A plot showing correlator offset for the High Band Antenna array of the LOFAR station at \nancay.  Note that the plot scale is 3~orders of magnitude smaller than the corresponding EMBRACE plot.  The outliers are all neighbour antennas indicating the offset is due to cross talk between those antennas.}
\label{fig:lofarcorroffset}
\end{figure}

The EMBRACE instrument has features in common with a LOFAR station. Their
backends are virtually identical allowing convenient comparisons of
data products. The LOFAR data normally show no or little sign of
correlator offset problems. By selecting correlations from 16 tiles
(LOFAR HBA) and plotting them in the
complex plane (Fig.~\ref{fig:lofarcorroffset}), we can see that for
LOFAR, at least 100 out of 120 unique baseline correlations are
concentrated symmetrically around the origin (within a 1-2 thermal
spreads when observing for 20 mintues). The 10-20 outlier clusters (on
the order of 10 thermal spreads from the origin) are all pairs who are
direct neighbours or two tiles/antennae away, possibly indicating that
the offset is due to crosstalk between array elements.  Overall we
conclude that the LOFAR correlations behaves fundamentally different
than for EMBRACE and more in line with what one should expect from an
ideal interferometer, namely that the expectation value of
correlations between array elements is zero when not observing a
source.
\refereedelete{(i.e. having random relative phases at a given instance in time).}

\subsection{Local oscillator distribution as a potential source of correlator offset}
\label{sec:LOcorroffset}
\begin{figure}[ht!]
 \centering
 \includegraphics[width=0.45\linewidth,clip]{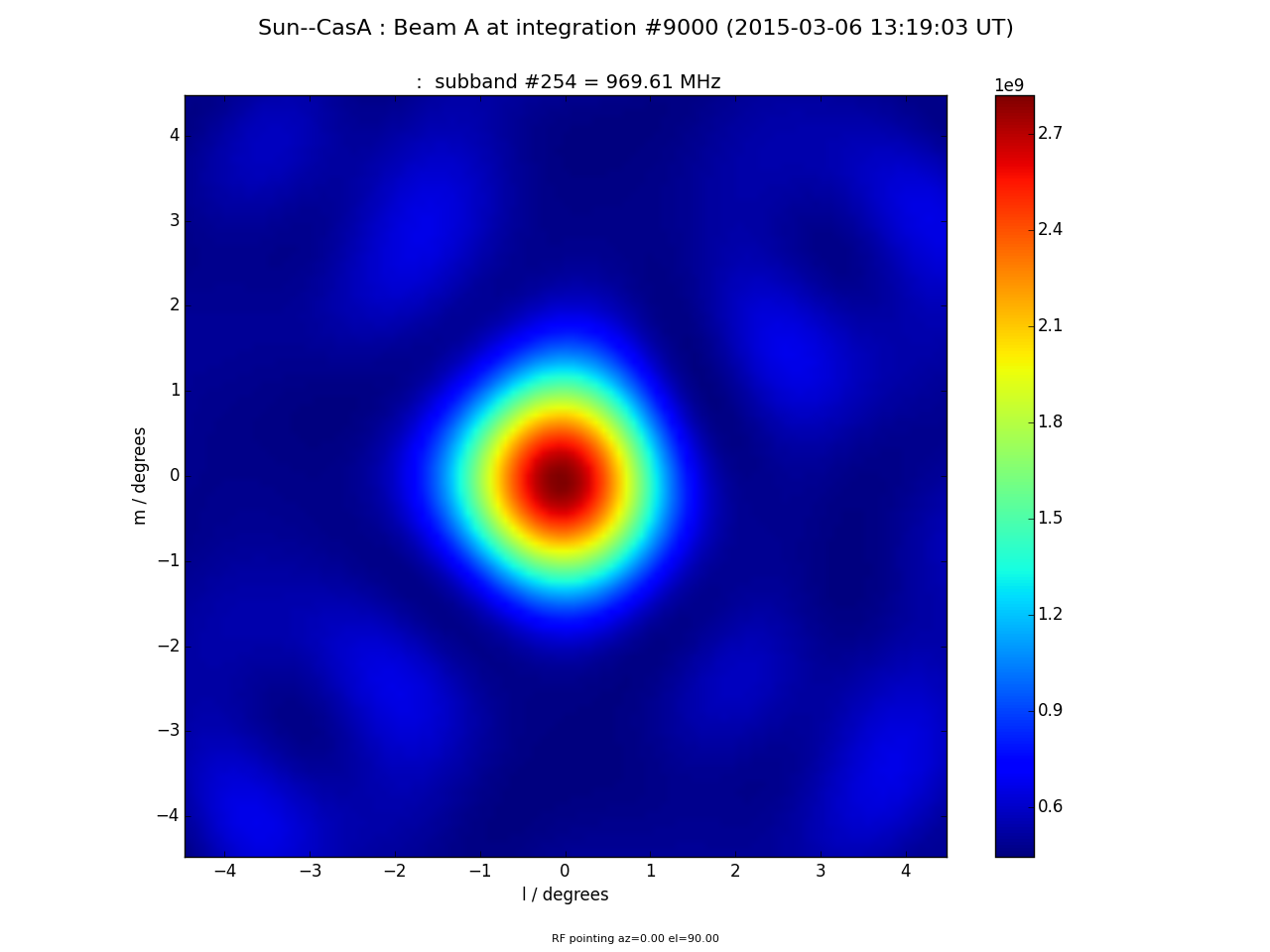}
 \includegraphics[width=0.45\linewidth,clip]{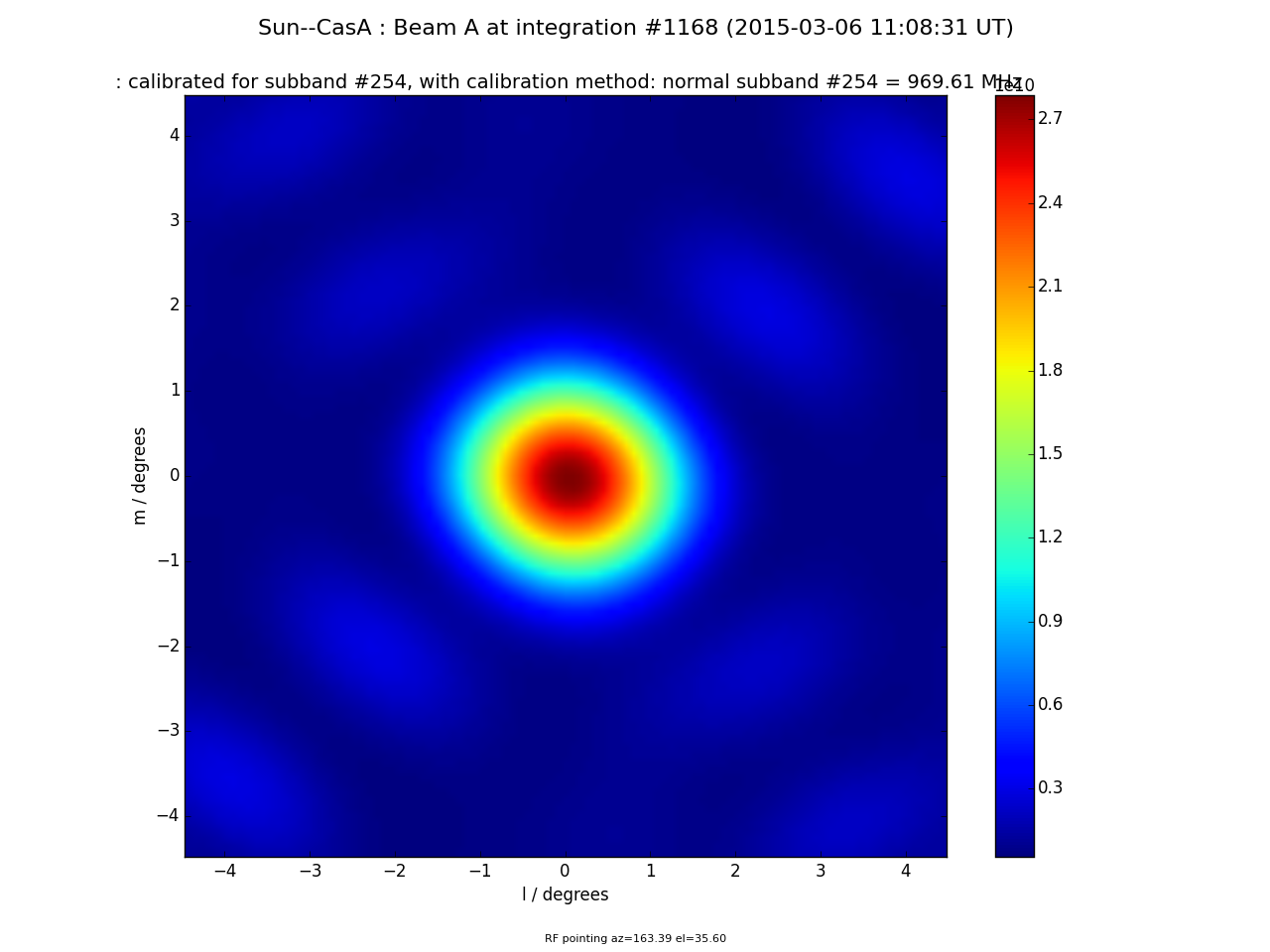} 
  \caption{Sky images without and with a source in the field.  Using the cross-correlation statistics data,
    interferometric images can be made at a given sub-band.  a)~~A
    phantom point source can be seen in images produced from data
    pointing at ``empty sky'' and assuming zenith pointing.  b)~~The
    image of the Sun has peak more than an order of magnitude larger
    than the phantom point source.}
  \label{fig:phantomPointSource}
\end{figure}

The correlator offset described above is seen as a constant offset in
the cross-correlations between the
tilesets of EMBRACE.  It behaves as a constant signal which is common
in all the tilesets and is correlated by the EMBRACE backend.  One can
see this by producing a skymap image in the usual manner while
assuming the array is pointing at zenith.
Figure~\ref{fig:phantomPointSource}a is an image of ``empty sky'' from
a drift scan of the Sun after the source has left the field of
view. There are no detectable sources in the field.  The result is
clearly an image of a point source, but with a much lower intensity
level compared to an image of the Sun (Fig.~\ref{fig:phantomPointSource}b).

The point source of Fig.~\ref{fig:phantomPointSource}a can be seen
at any time in the data as long as there is no strong source in the
field of view.  It is always the same, as already noted above, and is
independent of the instrument pointing parameters.  In order to create
this stable, ``phantom'' point source, there must be a signal common
to all the tilesets analogous to a source in the sky impinging a
signal on all the tilesets, but the signal must be internal to the
instrument and independent of pointing parameters.

The frequency downconversion stage described in Sect.~\ref{sec:CDC}
uses signal generators and power splitters to provide the Local
Oscillator (LO) signal to all the CDC cards.  This LO distribution is
an obvious candidate for the source of the phantom point source
(\ie~the correlator offset).  Although care has been taken to avoid
having LO or mixing products within the RF band, it appears that the
LO or mixing products are nevertheless modulating signals on the CDC
card, possibly by radiating onto the CDC card.

\section{Astronomical measurement of the embrace geometry}
The 72 Vivaldi antennas within a tile are laid out in rows rotated
45$^\circ$\ from the tile side so that the main diagonal has 12
elements separated by 12.5~cm. Then the 64 tiles in the array are
placed side by side in a right-angle rhombus pattern with respect to
the North-South meridian.  Distances within tiles and tilesets are
easily measured down to the necessary accuracy (fractions of a
wavelength), and it turns out that the nominal values can be used,
e.g. neighbouring tileset distances along a row can be assumed to be
2$\times$0.125$\times$12/$\sqrt{2}$\ metres apart.  Thus it is easy to
construct the 4$\times$4\ matrix of relative positions needed for the
complex weight calculation in the beam forming procedure, however,
such properties as overall tilt and rotation must also be known down
to fractions of a degree to avoid problems with pointing offsets when
tracking sources (compare with a minimum synthesized beam size of
$\sim$1.1$^\circ$).

The rotation was initially known within a few degrees which is
sufficient for shorter observations because the digital phase
calibration (see Sect.~\ref{sec:digproc}) will correct the error as
long as the pointing direction does not change much compared to the
position of the calibration source during the time of phase calibration.

In 2012, two independent efforts were made to more accurately measure the
rotation, both mechanically and by using a celestial source. In the latter
case we made use of the low cadence tileset cross correlations produced in the
backend every second and dumped into FITS files by the control software.
The Sun was tracked starting approximately mid-day (19~July~2012) from south
to west thus spanning almost a quarter of the local sky. An image was made
within the tilesets' FoV every 10 minutes and by retroactively assuming
different instrument rotations when computing the complex weights for the
phase calibration and actual data imaging, we could iteratively converge
towards the best number that kept the Sun's maximum intensity pixel at the
expected position throughout the measurement. The number arrived at with this
method was $\Delta\Theta$=-3.7$^\circ$. Incidentally this exercise was also an
important step in verifying that all sign and ordering conventions were
consistent between the various coordinate frames and cross correlation data
files.

The direct estimate of the physical rotation of the instrument was done by
measuring positions of corner antennas in three dimensions with respect to
a reference point at the observatory site. The resulting offset angles along
the north-south and east-west lines were found to be
$\Delta\Theta$=-3.673$^\circ$\ and $\Delta\Theta$=-3.714$^\circ$, respectively.
It should be noted that although these physical measurements were done about
one month prior to the on-sky experiment described above, the result was not
yet known to the observer during the data analysis.
The on-sky result thus turned out to be an excellent blind verification of
the geometric parameters of the instrument.

The tilt angle was found to be small, ca 0.06$^\circ$, mostly about the
north-south axis, estimated from a one centimetre height difference of two
reference points separated by nine metres along the east-west direction.

Those results have been confirmed by another method based on a model
point spread function fit on a point source skymap
(see Sect.~\ref{sec:beamsim}). This is a sampled rectangle aperture
model, with physical rotation and tilt of the array as 2 of the free
parameters to be optimized. The best fit performed on several
observations gives a consistent result of
$\Delta\Theta$=-3.714$^\circ\pm 0.0005$ for the rotation angle, and a
tilt angle ranging from $10^{-4}$ to a few $10^{-3}$ degrees.

\section{Beam pattern}
\label{sec:BeamPattern}
\subsection{Beam pattern measurements}
At the single tile level, the main beam was first seen to agree with
expectations for a one metre square phased up array when bore sight drift
scans of GPS L2 carriers were conducted in October/November 2009.
This result is described in detail in \citet{olofsson_limelette}.

The fundamental properties of the tileset beams and the full array beam
were quickly established after the installation of the hardware was finished
and phase calibration schemes were implemented during the summer months of 2011.

\begin{figure}
\centering
\includegraphics[width=0.9\linewidth,clip]{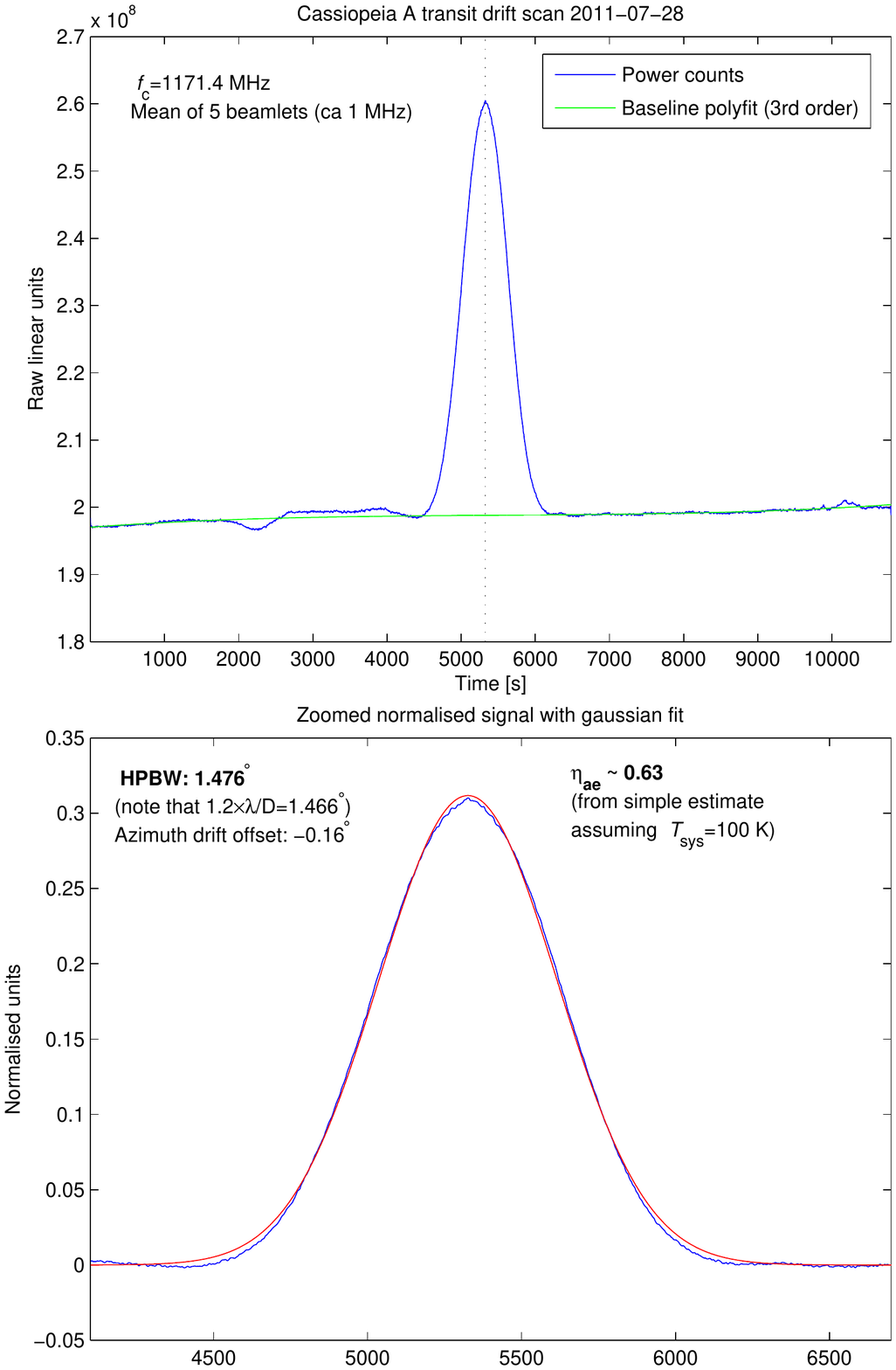}
\caption{
Drift scan of Cas~A at 1171.4~MHz in a bandwidth of $\sim1$~MHz.  The
top image shows the full time span of the drift scan with a baseline
fit.  The bottom plot is a zoom to the main lobe with the baseline fit
removed.  The measured FWHM is $1.476\deg$ which is very close to the
expected value of $1.466\deg$.}
\label{fig:casa}
\end{figure}

Figure~\ref{fig:casa} shows an example of a drift scan of Cassiopeia~A
pointing at its maximum elevation point. The east-west HPBW was
computed by converting the time axis to an angular scale and performing a
Gaussian fit to the normalized baseline subtracted signal.

Figure~\ref{fig:tsbeam} shows the time signals from a satellite carrier
drift scan for all individual 16 tile sets that constitute the array
members in the final digital beam forming. The slight dispersion of
the times of maximum signal is caused by the 45$^\circ$\ quantization
of the phase settings that are used to create (or `steer') the
analogue tile beams, and to measure the global phase offsets between
tiles in a tile set (see Sect.~\ref{sec:RFcal}).

\begin{figure}
\centering
\includegraphics[width=0.9\linewidth,clip]{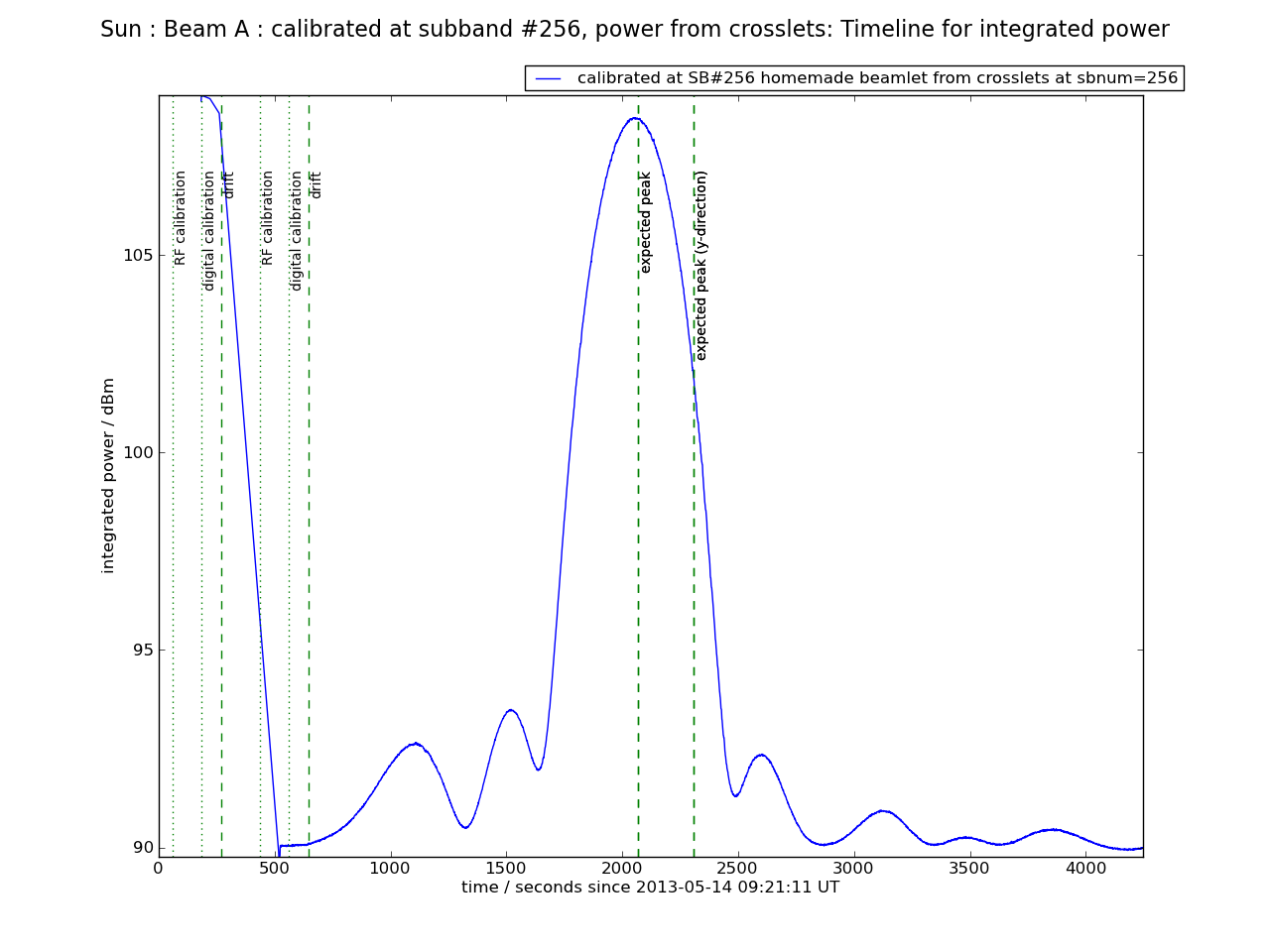}
\caption{
The drift scan of the Sun across the stationary digital beam pointing
reaches its maximum at the expected time.  The sidelobe levels are
below the 15~dB design criterion.
}
\label{fig:sundrift}
\end{figure}

Sidelobe levels (along the east-west direction) can be seen in the logarithmic
intensity scale solar drift scan in Fig.~\ref{fig:sundrift}. The first
secondary lobes can be seen to be $\gtrsim$15~dB down from the main lobe
which is within design goals.

\begin{figure}
\centering
\includegraphics[width=0.95\linewidth,clip]{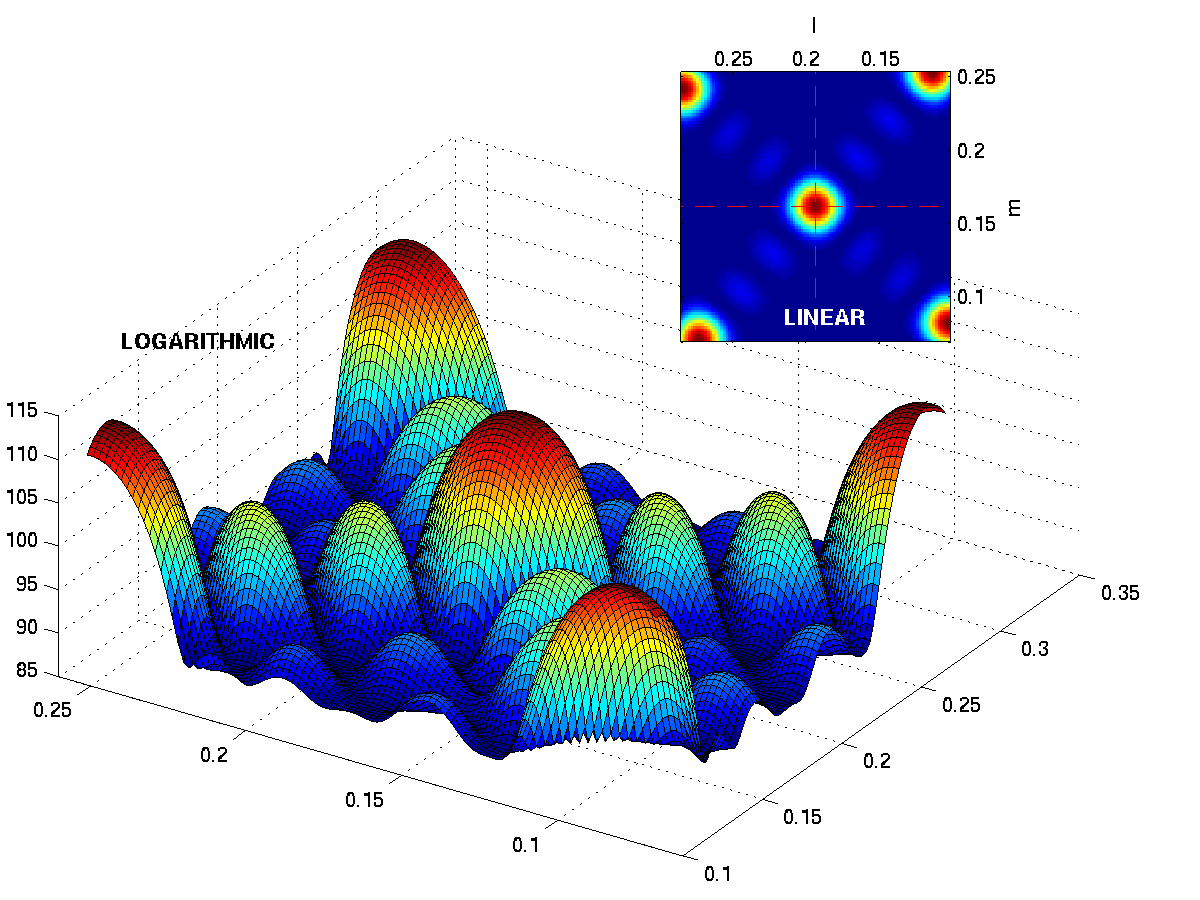}
\caption{Snapshot image of the GPS BIIF-1 satellite on 2012-09-05 at 1175~MHz.
The angular range covered approximately corresponds to two times the tileset
FoV. Note that for the full up instrument beam layout, the pattern shown is
tapered by the larger scale tileset main beam which is outlined in
Fig.~\ref{fig:tsbeam}.}
\label{fig:gpsim}
\end{figure}

The 2D sidelobe pattern at the smallest spatial scales could
conveniently be studied once algorithms were in place to read and
analyse the raw cross correlations data product.  Figure~\ref{fig:gpsim}
shows a narrow band instantaneous radio image of a point source (GPS
satellite) in linear and logarithmic units.  This can be compared with
the simulated pattern (see Sect.~\ref{sec:beamsim}). Note that the sky
projection used (collapsing the azimuth/elevation sky position onto a
flat plane by taking the cosine of the elevation) falsely leads to a
beam shape that is independent of elevation, whereas on the sky the
beam gets elongated in the up-down dimension at low elevations due to
the shortening projected baselines (zero at the horizon). The pattern
is repeated along the NW-SE and NE-SW axes due to the symmetry of the
tileset grid and the resulting ambiguity that arises when
phase-shifting exactly one whole wavelength on the shortest baselines
(2.1~m) to create the image pixels.

\subsection{Beam pattern simulation}
\label{sec:beamsim}
When pointing towards an unresolved source, the cross correlation
statistics data (once per second, per sub-band) gives access to a full
2D measurement of the Point Spread Function (PSF) of the instrument.

We fit a sampled rectangle aperture model to the cross correlation data with
array length dimensions $a$ and $b$, rotation and tilt angles of the
array respectively $rot$ and $tilt$, field of view width $fov$, and power
offset $P0$ as free parameters, following the theoretical square
aperture:
\begin{equation}
PSF(\theta,\phi)=(1-P0)\left[sinc^{2}(\frac{k\theta a}{2\pi})\, sinc^{2}(\frac{k\phi b}{2\pi})\right]+P0
\label{eq:PSF}
\end{equation}
where $\theta$ and $\phi$ are angles with respect to the
centre of the field of view, and taken along two great circles on sky
that have been rotated by an angle $rot$ and tilted by an angle $tilt$
with respect to the celestial meridian.

When periodized, this model may be seen as an ideal case where the
array would behave exactly as a sampled square aperture with
independent and identical receivers on its whole surface. It is thus
taking limitations such as diffraction-limited resolution, overall
beam shape, or PSF aliasing into account, but neglects other possible
influences like gain variations across the array or mutual
coupling. As already discussed in Sect.~\ref{sec:redcal}, those latter
effects could nevertheless explain some fine effects, and the gain variations are corrected by the redundancy method.

\refereedelete{While complex gain variations between tilesets are corrected by
applying the redundancy based calibration, mutual coupling is thus
expected to explain a significant part of the residuals when comparing
the observed PSF and the model, which opens a possibility for
characterizing those couplings.}

\begin{figure}[ht!]
\centering
\includegraphics[width=0.9\linewidth,clip]{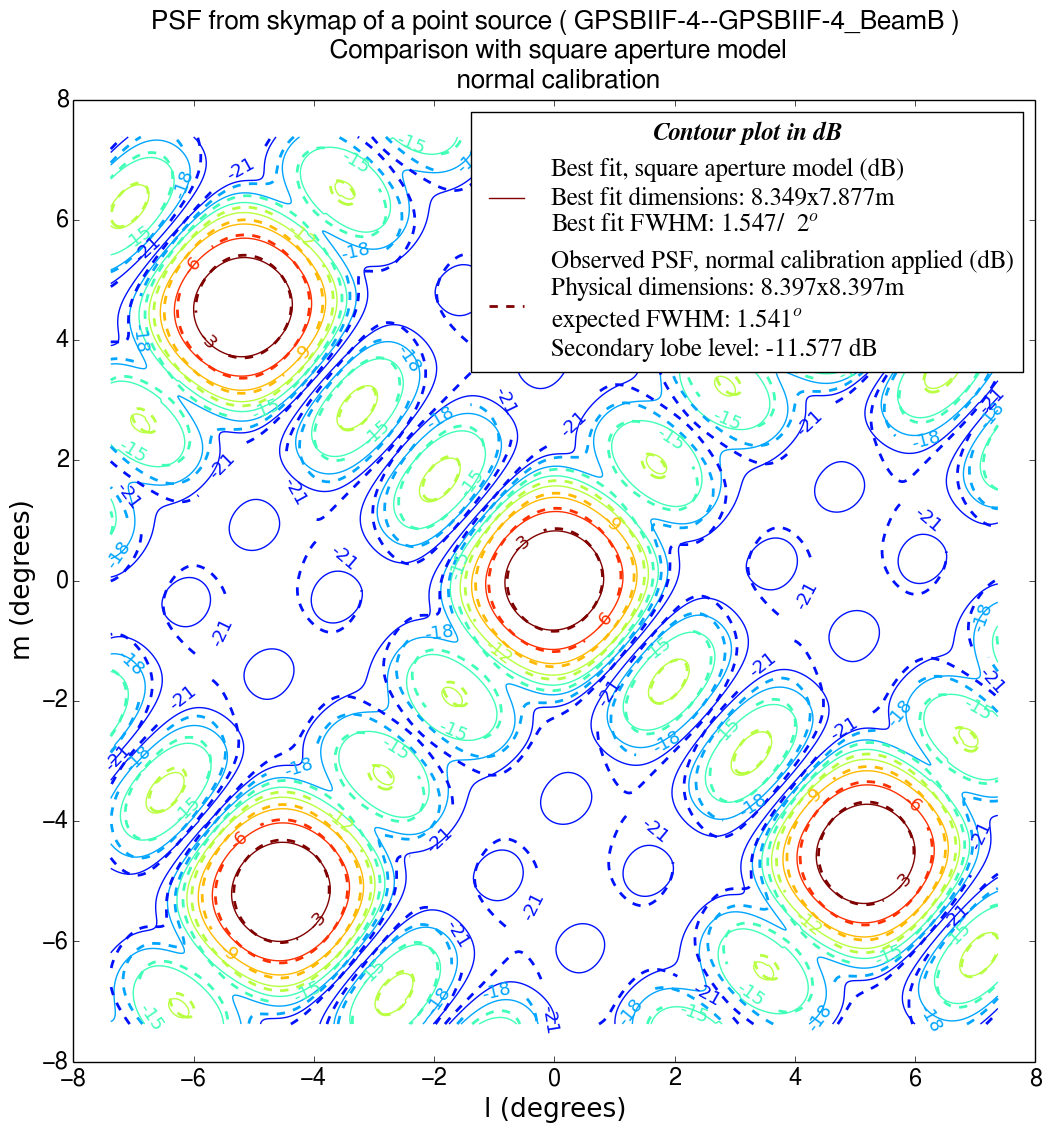} 
 \includegraphics[width=0.9\linewidth,clip]{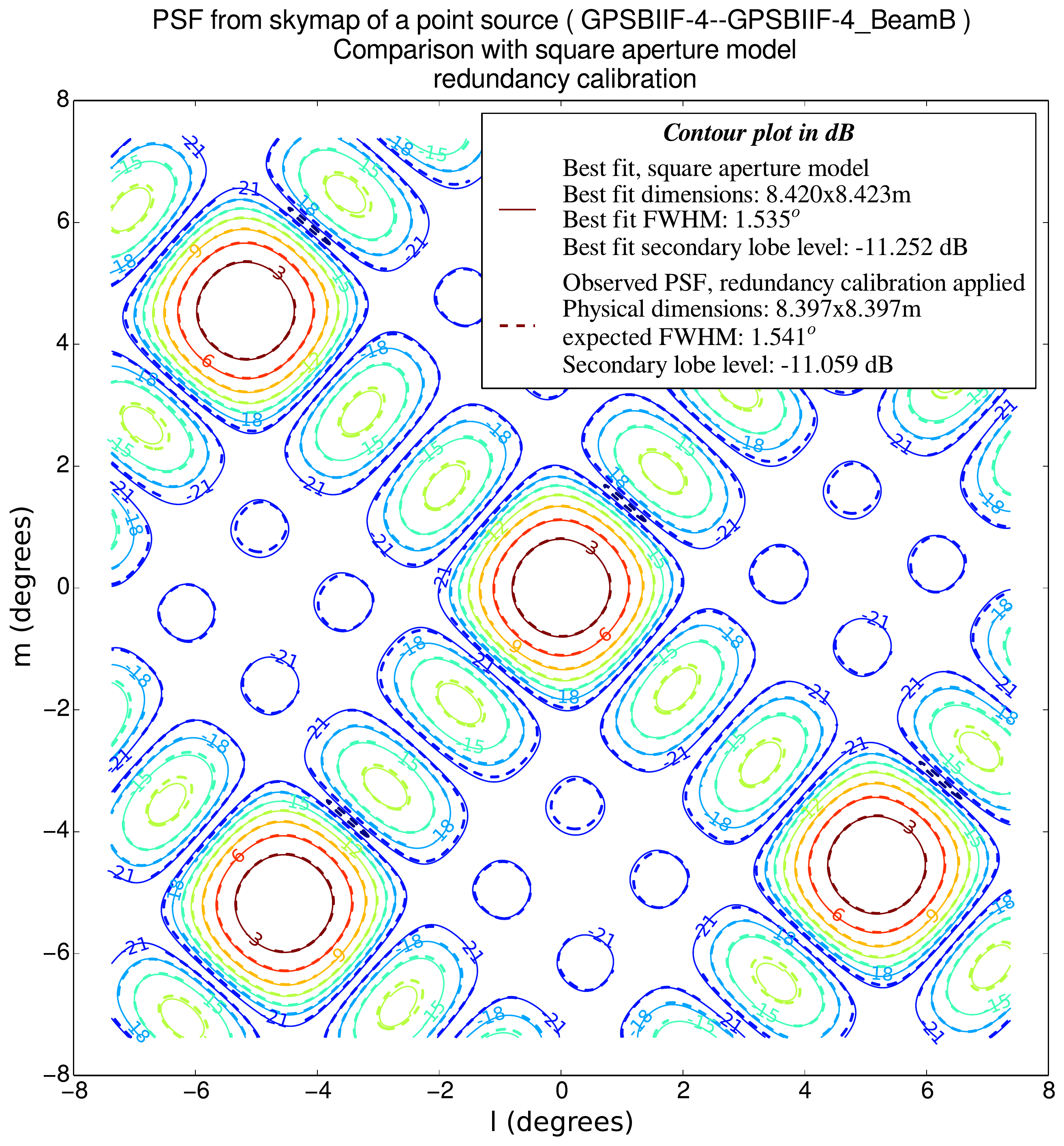}
  \caption{Contour plot (dB scale, 0 at maximum amplitude) of the
    interferometric imaging in one sub-band and after a phase-only
    calibration (top panel), and a full redundancy-based calibration
    (bottom panel), for a GPS satellite observation with \enancay.
    Observed data are shown as dashed contours, while the best-fit
    rectangle aperture model is shown as continuous ones.  The
    periodization of the main pattern is due to sampling of the UV
    plane, here the equivalent of 3~fields of view is shown.}
  \label{fig:PSF_contour}
\end{figure}

Figure~\ref{fig:PSF_contour} shows both observed PSF and a best-fit
as contour plots of an interferometric image in one sub-band, during
the observation of a GPS satellite. Two cases are presented: one with
phase-only calibration, and one where the complex gains of each
tileset has been corrected by redundancy-based calibration on the same
source. The root mean square residual amplitude between redundancy
calibrated data and best-fit model is less than 0.3\%, and about twice
as large in the phase-only calibration case.

\begin{figure}[ht!]
\centering
\includegraphics[width=1\linewidth,clip]{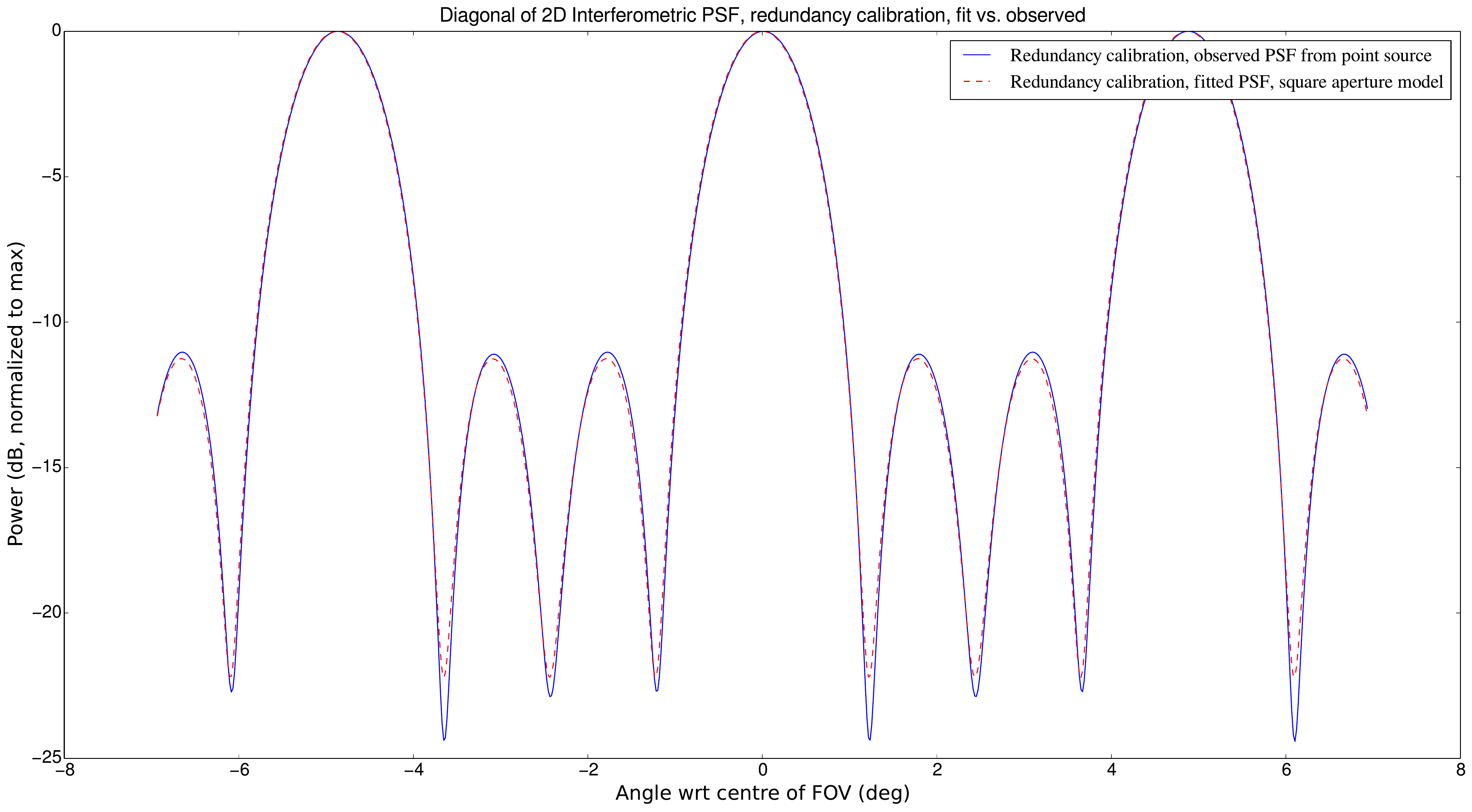} 
  \caption{1D Diagonal of 2D interferometric PSF (dB scale, 0 at
    maximum amplitude), for both GPS observation and best-fit model.}
  \label{fig:1D_PSF_diag}
\end{figure}

One may note the periodization of an underlying PSF pattern
corresponding to a perfectly sampled UV plane. Since our sampling is
here limited to tilesets, and given the total size of the array, this
periodization leads to an aliasing effect that tends to increase the
Side Lobe Level (SLL). Our model takes it into account by periodizing
the expression given in Eq.~\ref{eq:PSF}, with an angular period
corresponding to the field of view $\frac{\lambda}{d_{min}}$, where
$\lambda$ is the observed wavelength and $d_{min}$ is the minimum
separation between two tilesets. As a result, the observed SLL is
bigger than the expected one for a fully sampled rectangular aperture
(-13.2~dB of expected attenuation for the first side lobe
\citep{antenna_handbook}, -11.9~dB when summing the first and second
side lobes), as seen in Fig.~\ref{fig:1D_PSF_diag}.

\begin{figure}[ht!]
\centering
\includegraphics[width=0.9\linewidth,clip]{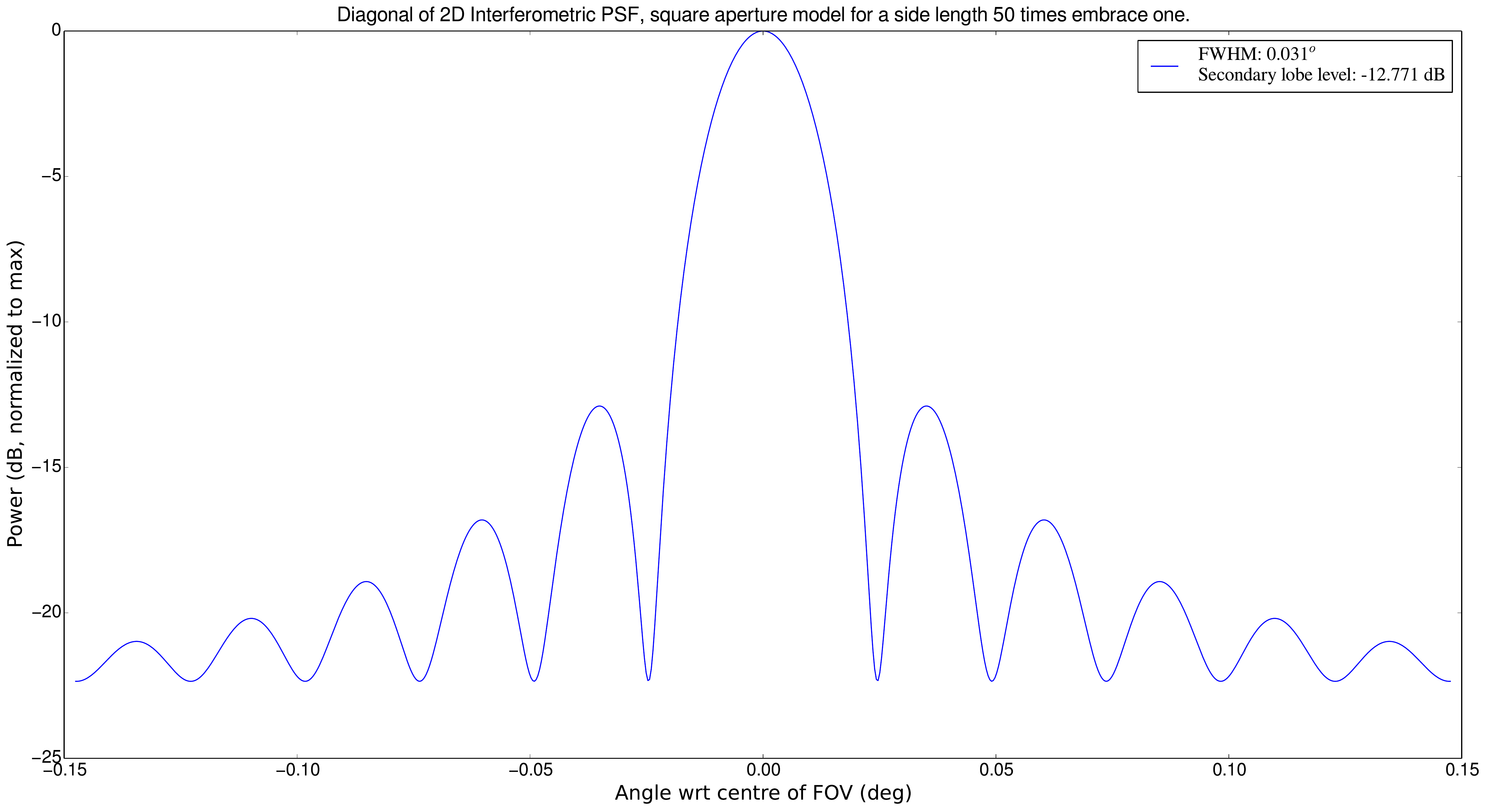} 

  \caption{1D Diagonal of 2D interferometric PSF (dB scale, 0 at maximum amplitude), for an EMBRACE-like array with 50~times larger side length (about 400~meters side length), following the square aperture model best-fit parameters.}
  \label{fig:1D_PSF_bigembrace}
\end{figure}

This limitation is expected, given the relative small size
of \enancay. We may extrapolate our model for a larger array, with
higher spatial resolution and thus less of these aliasing
effects.  Figure~\ref{fig:1D_PSF_bigembrace} shows the modeled PSF for
an EMBRACE-type station with 50~times longer sides (2500~times bigger
area). The remaining aliasing contribution to the first side lobe
level is of the order of 0.4~dB.

After redundancy calibration, the observed Full Width Half Maximum
(FWHM) is smaller than theoretically expected, reflecting the
over-estimated best-fit length and width. By comparison, the phase
calibrated data (no gain amplitude correction between tilesets) shows
an asymmetric rectangular aperture model, with one side length as
expected and another one under estimated, reflecting the degraded
spatial resolution in this direction along the side. The behaviour of
tileset amplitude gain corrections given by the redundancy method
shows this can be interpreted in terms of an apodization effect. For
these observations, the tileset gains repartition over the array is
indeed close to a bilinear function, with a gradient aligned with one
side of the square array. Before gain amplitude correction, the array
thus behaves with a smaller effective area, and a degraded angular
resolution in one direction. This kind of regular pattern in gain
correction over the array's tilesets is sometimes encountered, but it
may vary from observation to observation. The origin is not yet fully
understood, but the redundancy calibration can correct for the gain
variation. Starting from a corrected array, one can then choose a
tileset apodization scheme (complex weights) that will suit the
desired trade-off between SLL and FWHM.

On the other hand, the best-fit SLL on fully calibrated data is also
bigger than the expected one for a realistically sampled square
aperture by about 0.1~dB (see Fig.~\ref{fig:PSF_contour} and Fig.~\ref{fig:1D_PSF_diag}). This residual may be interpreted in terms
of mutual coupling, and a simple toy model allows an estimate of the strength
of these couplings between closest elements on the order of a few percent at most.
That interpretation will be further investigated. If the coupling origin is confirmed, a more precise quantification at different frequencies and line of sight as well as the possibility for correcting it will be tested \citep{myarticle_mutual_coupling}.

\section{Multibeaming}

\begin{figure}[ht!]
 \centering
 \includegraphics[width=0.9\linewidth,clip]{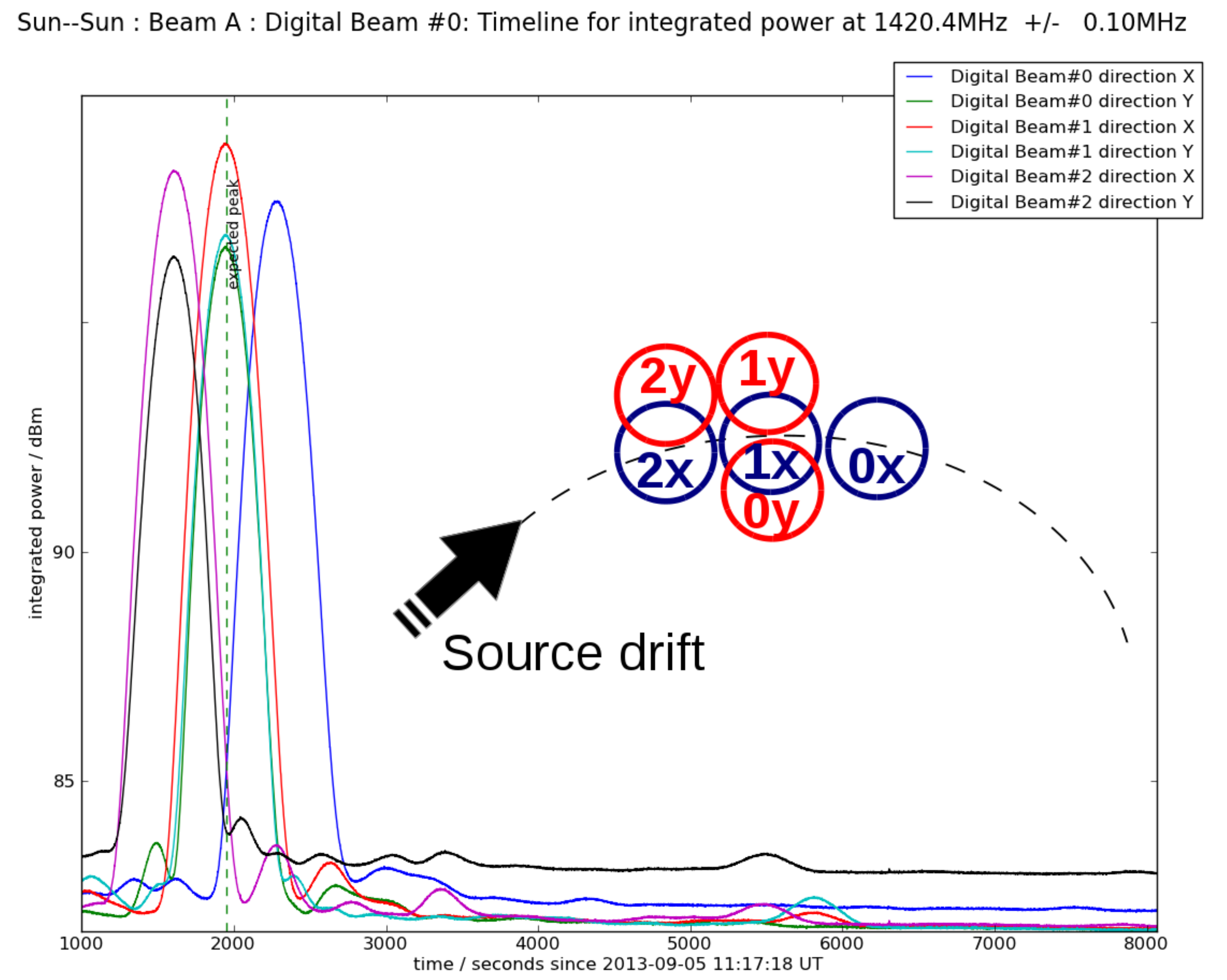}      
  \caption{This drift scan of the Sun used 6~beams of
    \enancay\ pointing along the trajectory of the Sun across the sky
    (inset top right).}
  \label{fig:multidrift}
\end{figure}

Figure~\ref{fig:multidrift} shows a drift scan of the Sun using the
multibeam capability of \enancay.  Six beams were pointed on the sky
along the trajectory of the Sun, including three partially overlapping
beams.  The result shows the Sun entering and exiting each beam and
the off-pointed beams are approximately 3~dB down from the peak, as expected.

\section{Noise performance and efficiencies}
\subsection{Sensitivity/efficiency}
\label{sec:efficiency}
\newcommand{\Aphys}{\mbox{$A_{\rm phys}$}}
\newcommand{\Aeff}{\mbox{$A_{\rm eff}$}}

The system noise cannot be directly determined without filling the whole beam
pattern with two known noise sources (such as cold sky and a hot load that
covers the entire array) which is not practical for an instrument such
as EMBRACE, however, the \emph{ratio} of noise and efficiency can be determined while
observing small sources with known flux densities or known brightness
temperature distributions.

For instance, the expected main beam brightness temperature from the Sun is
\begin{equation}
T_{\rm mb}=T_{\rm b}^{\rm Sun}\times\eta_{\rm bf}
\end{equation}
where $\eta_{\rm bf}$\ is the beam filling factor for a disk-shaped source
of size $\theta_{\rm S}$\ and a Gaussian HPBW main lobe $\theta_{\rm B}$:
\begin{equation}
\eta_{\rm bf}=1-e^{-ln2\,\theta_{\rm S}^2/\theta_{\rm B}^2}
\end{equation}
Given a linear receiver in the Rayleigh-Jeans regime and a switched
measurement with total power values (or spectra)
`on' and `off', the antenna temperature should be
\begin{equation}
T_{\rm A}=T_{\rm sys}\mathit{\frac{on-off}{off}}
\end{equation}
if we neglect the CMB and atmospheric contributions which are small in this
context. By using the relation \mbox{$T_{\rm A}=\eta_{\rm mb}T_{mb}$} we can now
put everything together to find a simple expression for the ratio of system
temperature and main beam efficiency:
\begin{equation}
\frac{T_{\rm sys}}{\eta_{\rm mb}}=\frac{T_{\rm b}^{\rm Sun}\eta_{\rm bf}}{N_{\rm p}}
\end{equation}
where $N_{\rm p}$\ is shorthand for the normalized raw power (on-off)/off.

Unfortunately, the Sun is highly variable at decimeter wavelengths and longer
where the chromosphere and later the corona is starting to dominate over the
photosphere. There are no telescopes that
monitor the Sun in the low L-band range on a daily basis and we are forced
to use interpolations for ``Quiet Sun'' brightness temperatures and accept
that the result may be uncertain by up to $\sim$50\%.

For small sources with relatively well known flux densities, one can more
directly arrive at another performance measure, namely the ratio of system
noise and aperture efficiency. The latter is simply the effective
collecting (\Aeff) area over the physical area (\Aphys), which can be measured
by comparing observed power from a source with its intrinsically
available power over the instrument as given by the flux density
multiplied by the bandwidth.
\refereedelete{
\begin{equation}
P_{\rm source}=\Aphys\frac{S_{\rm source}}{2}B
\end{equation}
and
\begin{equation}
P_{\rm obs}=\Aeff\frac{S_{\rm source}}{2}B=\eta_{\rm a}\Aphys\frac{S_{\rm source}}{2}B
\end{equation}
The source flux density is divided by half since the instrument only picks
up one polarization component (we assume the source is unpolarized).
Similarly to the expression for the antenna temperature above for a switched
measurement, we can also express the detected power as
\begin{equation}
P_{\rm obs}=(kT_{\rm sys}B)\mathit{\frac{on-off}{off}}
\end{equation}
Combining the two expressions for $P_{\rm obs}$\ we can write
}
The resulting expression is
\begin{equation}
  \label{eq:sefd}
S_{\rm source}=\frac{2kT_{\rm sys}}{\eta_{\rm a}\Aphys}N_{\rm p}
\end{equation}
where $T_{\rm sys}$/$\eta_{\rm a}$\ is the sought quantity.  Often the entire fraction in Eq.~\ref{eq:sefd} is used instead:
\[
\frac{2kT_{\rm sys}}{\eta_{\rm a}\Aphys}\triangleq{SEFD}
\]
where SEFD refers to ``System Equivalent Flux Density''.

An additional complication in the case of a dense aperture array is that the
cross sectional area goes down for elevations progressively away from
zenith and this can be viewed either as a decrease in physical area, or
as a decrease in $\eta_{\rm a}$. Theoretically, if each Vivaldi element
was truly isotropic and the array perfectly dense, the received power
should decrease proportionally to the projected area, i.e., as
$\sin(el.)\times{}\Aphys$.  In reality it does not follow
such a curve and it is of interest to try to characterize its elevation
dependence. Since the beam pattern is not circular symmetric, the preferred
manner to do this would be to track a source along fixed azimuths from
zenith to low elevation. No sources follow such sky trajectories\footnote{It's worth noting that a satellite in Low Earth Orbit with an isotropic
transmitter would be a rather good approximation due to its swift passage
across the sky.} and we will have to draw tentative conclusions based on
results interleaved between our stronger sources (Sun, GPS, Cas~A, Cyg~A).

Lastly, it can also be of interest to compare the performance between the
different hierarchical stages in the interferometer. Here, we restrict
ourselves to compare the area efficiency of a single tileset to that of the
whole array.

As can be seen, there is a difference of approximately a factor of two
and we could attribute this to an additional `array efficiency' as defined
in \citet{interfero_bible}. Using that formalism, the $\eta_{\rm a}$\ used previously
for the full array would then be replaced with $\eta_{\rm a}\eta_{\rm array}$\
where $\eta_{\rm a}$\ now is the tileset value.

\subsection{Elevation limit due to the forest}
\label{sec:treeline}
\begin{figure}
\centering
\includegraphics[width=\linewidth,clip]{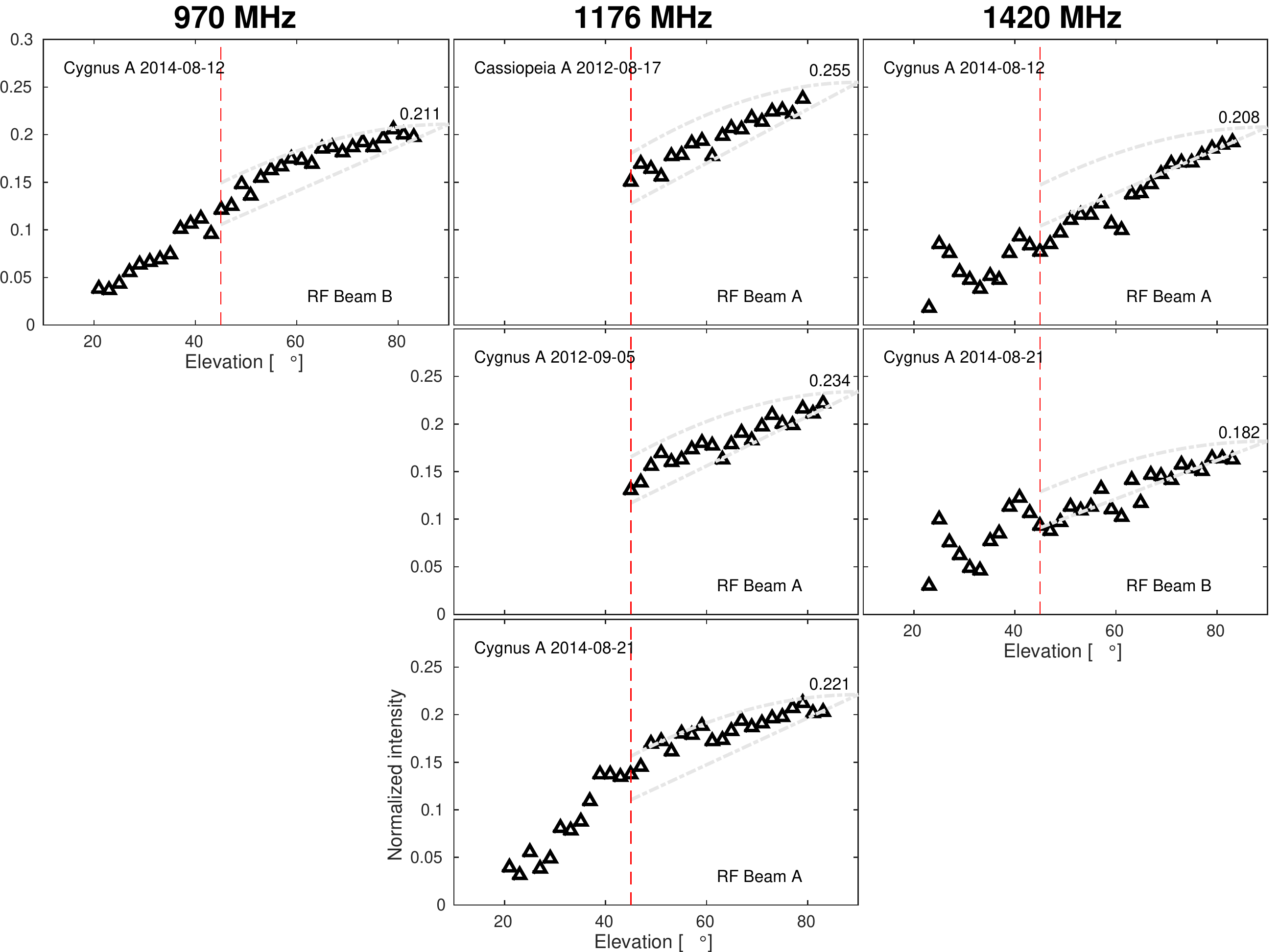}
\caption{Received intensity as a function of elevation for CasA
  and CygA at three frequencies.  \modified{The vertical red lines indicate
  45\degrees\ elevation which is the nominal scan angle limitation for
  EMBRACE.}  The effect of the treeline is also visible at elevations
  below 45\degrees.}
\label{fig:treeline}
\end{figure}

Figure~\ref{fig:treeline} shows the normalized intensity for a number
of observations as a function of elevation at three frequencies, using
either of the two RF beams. The trends at high elevations seem mostly
linear and we have tried to extrapolate these trends in order to
phenomenologically estimate the value at zenith (noted at the right edge
of the plots).

As comparisons to the actual measured values, we have drawn two
lines. The upper dashed line is the trend the received power would
follow at lower elevations if the instrument was a perfectly flat,
fully sampled, infinitely thin, isotropic 2d array (i.e., having a
geometric cross section fall-off as the cosine of elevation). The
lower dashed line is simply a 1st order polynomial that goes through
the estimated peak value, and the origin (i.e., assuming zero power at
zero elevation).

When analysing the Cygnus A measurements, we consistently noted that
results appear more erratic below a certain elevation when the source
is setting. We have previously mostly only observed at low elevations
in the south and a theory was advanced that obstruction from local
vegetation could create the observed effects.  An inspection of the
treeline around the instrument confirmed that trees in the north west
block the view up to elevations of circa 45\degrees, whereas the view
to the north east -- where the source is rising -- is clear down to
much lower elevations.

\refereedelete{
We also note a briefer irregularity around 60\degrees\ elevation that
seems to be more pronounced at the highest frequency. From the raw
data, we can see that this is due to slightly nonsimultaneous steps in
the on- and off-beam powers typically leading to short-duration dips
in the normalized power curves, see e.g. the top right panel in
Fig.~\ref{fig:treeline}.

In general we note that the slopes (and therefore the effective area
as a function of elevation) are somewhat different at the three
frequencies investigated (using the reasonable assumption that the
system noise stays largely the same). The steepest slope/fall-off is
seen at the highest frequency. This could be a consequence of the
smaller beams of the individual Vivaldi elements at shorter
wavelengths. The divergence from the ideal isotropic (uniform) beam
pattern is thus larger the higher the frequency for a given off-zenith
direction.
}

\subsection{Stability}

It is already clear that the stability is such that we can successfully
subtract data to get rid of systemic patterns (see Sect.~\ref{sec:embdat} and
Sect.~\ref{sec:embspec}). We have also tried to estimate the Allan variance
although we cannot shield the instrument from the outside world.
Preferably one would like to expose the instrument to a constant signal
source such as a hot load. The results achieved, where we clearly still have
some variability due to external sources, should thus be considered a conservative
limit to the true Allan variance.

\begin{figure}
\centering
\includegraphics[width=0.9\linewidth,clip]{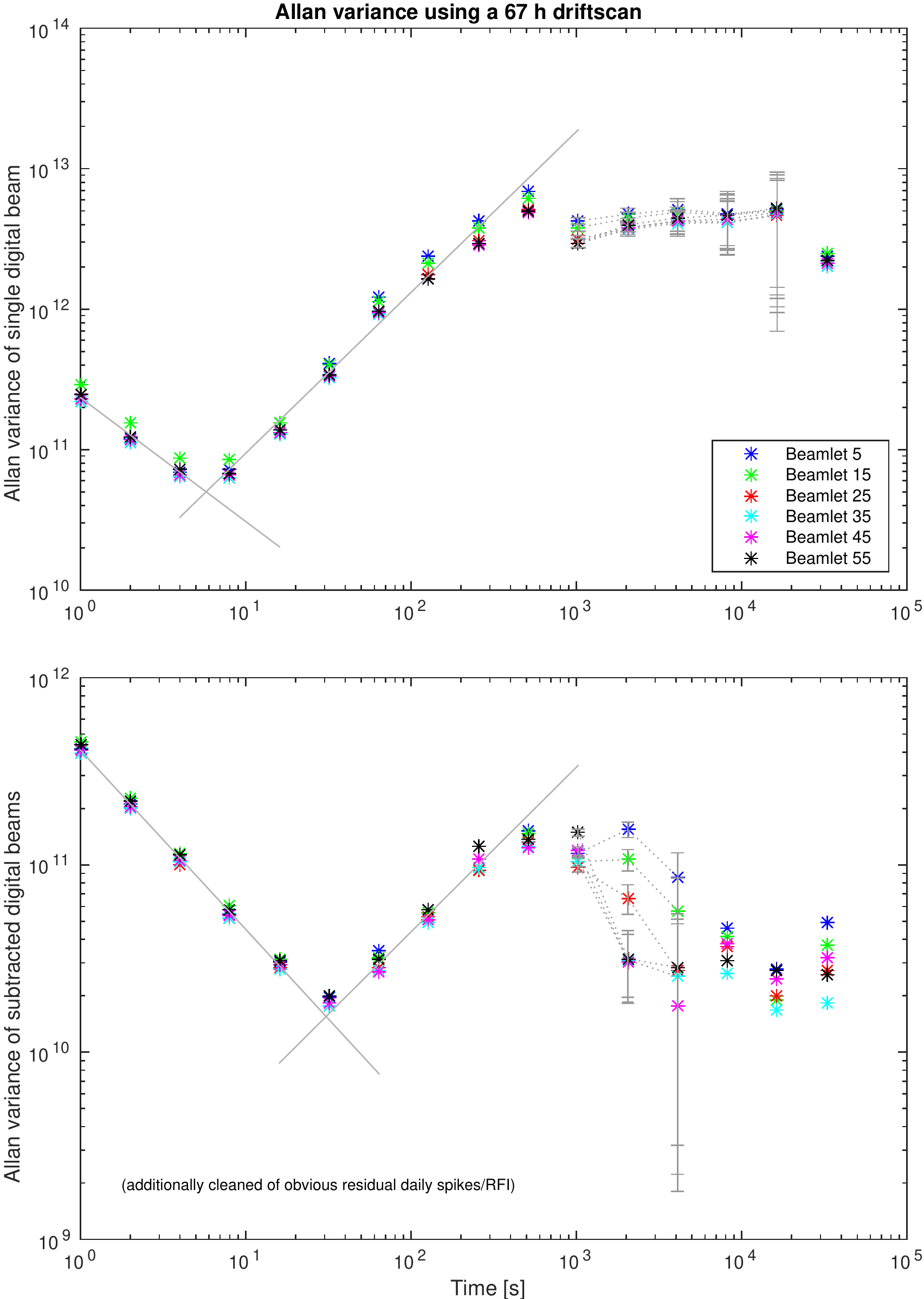}
\caption{Allan variance as a function of time. This observation was performed
as a three-day drift scan of Cas~A pointing at $\sim79^\circ$\ elevation due
south. The signals from the daily Cas~A passages have been removed. In the
lower panels also a few strong signals have been removed that occurred on a
solar day basis (those could be due to any kind of human activity such
as a transmitter that is scheduled to turn on or increase its power for a
short while every day, or artificially induced changes to the instrument
environment such as variations in the AC power grid).}
\label{fig:allan}
\end{figure}

In Fig.~\ref{fig:allan}, we show the Allan variance for a constant pointing
stretching over three days where the total power was sampled every second.
Values for six different beamlets are shown from a bandpass of 61 beamlets
(beamlet=sub-band here since all beamlets were pointing in the same direction). 
\modified{The lower panel shows the same type of analysis but for the
difference signal of two digital beams. This could be
considered a measure of what is sometimes referred to as the
``spectroscopic'' Allan variance which normally shows
better stability than the raw total power signal.
To further improve the lower limit to the best-case Allan
variance time we also} removed a small number of relatively narrow
strong features from the time signal. These signals occured with a period
of exactly one solar day and must hence be of terrestrial origin and as can
be seen they had an effect on the Allan variance down to time scales
$<$10~s. The lower panel demonstrates with some confidence that the
Allan time of the system (where drift noise starts to dominate over thermal
noise) is better than 30 seconds.

\section{Astronomical Sources}
\subsection{Cassiopeia A}
\label{sec:CasA}
\object{Cas~A} is one of our two good candidates for absolute efficiency
estimations due to its constant flux (on short time-scales) and being
a point source for EMBRACE. A detailed flux model over frequency and time for \object{Cas~A}
(which is an expanding supernova remnant that fades over time in the
radio) has not been presented since \citet{1977A&A....61...99B} who
assembled observations during the 1950's, 60's, and 70's and
calculated fading rates over a large set of frequencies and
extrapolated an equivalent flux density spectrum for the epoch 1980.
The 1980 values are still sometimes used today and they severely
overestimate the current flux density of \object{Cas~A}. On the other
hand, if one uses the fading rates of \citet{1977A&A....61...99B},
valid some 50 years ago, to extrapolate from 1980 to a much later
time, one significantly \emph{under} estimates the flux because there
is ample proof that the fading rates have decreased.

In our observations of \object{Cas~A} and \object{Cyg~A} in 2012, 
we noted that \object{Cas~A} seemed to be $\sim$10\% \emph{stronger} than \object{Cyg~A} at 1176~MHz. {This assumes
that the system and ambient noises were constant between the measurements
which may not be true but we find it unlikely that it varied by more
than 10\%.}  According to the flux model by \citet{1977A&A....61...99B}, the flux
for epoch 2012.6 at this frequency should be \emph{lower} than that of \object{Cyg~A}.
Since no large span absolute flux spectrum of \object{Cas~A} has been measured in
recent years, our approach to this conflict is to still use the 1980
model spectrum from \citet{1977A&A....61...99B}, but extrapolate using the few more recent estimates for fading
rates that can be found in the literature, notably at 927~MHz \citep{2004astro.ph.12593V}
and 1405~MHz \citep{2000ApJ...537..904R}, and made linear interpolations
to find fading rates for the frequencies observed with \enancay.
Although it does not perfectly fit with our observed
$S_{\rm CasA}/S_{\rm CygA}$, it does agree better than either using the Baars
model propagated to the 2010's or than directly using the 1980 spectrum
ignoring any fading during the last 30~years.

It should be noted that, according to \citet{2000ApJ...537..904R},
fading rate decrease is only seen at lower frequencies
(the transition being anywhere between $\lambda$$\approx$4-10~cm).
Furthermore, according to them, the fading rate of 0.6-0.7\% per year
previously seen at 7.8~GHz and above, now appears to apply to
\emph{all} frequencies down to a few 10ths of~MHz.

\subsection{Spectral line observation of \object{Messier 33}}
\label{sec:embspec}

With a relatively modest collecting area, strong distinct spectral signatures
are restricted to artificial sources such as satellite carriers, and the
Milky Way HI line at 1420.4~MHz. The latter is distributed throughout the
entire sky and mostly lacks clear unique features on spatial scales of the
array beam. An example of an early EMBRACE HI observation -- at a time when
tilesets could not yet be phased up to form an array beam -- was to point all
tileset beams to a certain longitude in the galactic plane and verify
that the resulting spectra had the Doppler components expected in that
slot of the plane compared to published surveys.

\label{sec:M33}
The Triangulum galaxy (\object{M33}) is the most distant object that can be observed with the naked eye
in the optical (under very good conditions). It is an ideal source to
observe with EMBRACE due to its size (about 1$^{\circ}$ diameter,
i.e., slightly smaller than the synthesized beam) and its radial
velocity ($v_{\rm LSR}$$\approx$-200~km$\,$s$^{-1}$).

For atomic hydrogen 21~cm observations (the strongest astronomical
spectral line within the EMBRACE bandpass by orders of magnitude), the
latter fact is very fortuitous; the line-of-sight orientation of the
\object{M33} disk leads to a relatively narrow line profile
(FWZI$\sim$200~km$\,$s) that is clearly separated from the much
stronger local Milky Way foreground which lie at at an LSR velocity
near zero in the direction of \object{M33}.  Most other nearby spiral galaxies
overlap with the galactic foreground and are weaker due to larger
distances. The nearby \object{M31} (Andromeda) overlaps marginally with local
gas but is much larger and closer to edge-on orientation and would be
more suitable for one-dimensional mapping observations with EMBRACE.

During the summer of 2013, \object{M33} was observed both at the coarse sub-band spectral
resolution (1~sub-band=195~kHz=41.2~km$\,$s$^{-1}$) as a prepatory test, and
then at high resolution by capturing the stream of UDP packets for half an hour
(160 Gbyte in total) and channelizing each sub-band 16-fold.
The two digital beams that are always present in the backend output
(referred to as the ``X'' and ``Y'' directions due to its LOFAR heritage)
were employed to create FoV symmetric on- and off-beams in a manner described
in Fig.~\ref{fig:obsstrat}.
A spectrum was subsequently created in the simplest way possible; by
coaveraging all data for each position separately before off-position
subtraction and normalization, and then multiplying by a constant factor
so that our result matches a template spectrum we retrieved from the
LAB survey~\citep{2005A&A...440..775K}\footnote{Via the web interface at
\url{http://www.astro.uni-bonn.de/hisurvey/profile/}} convolved to the
EMBRACE array beam size.

\begin{figure}
\centering
\includegraphics[width=\linewidth,clip]{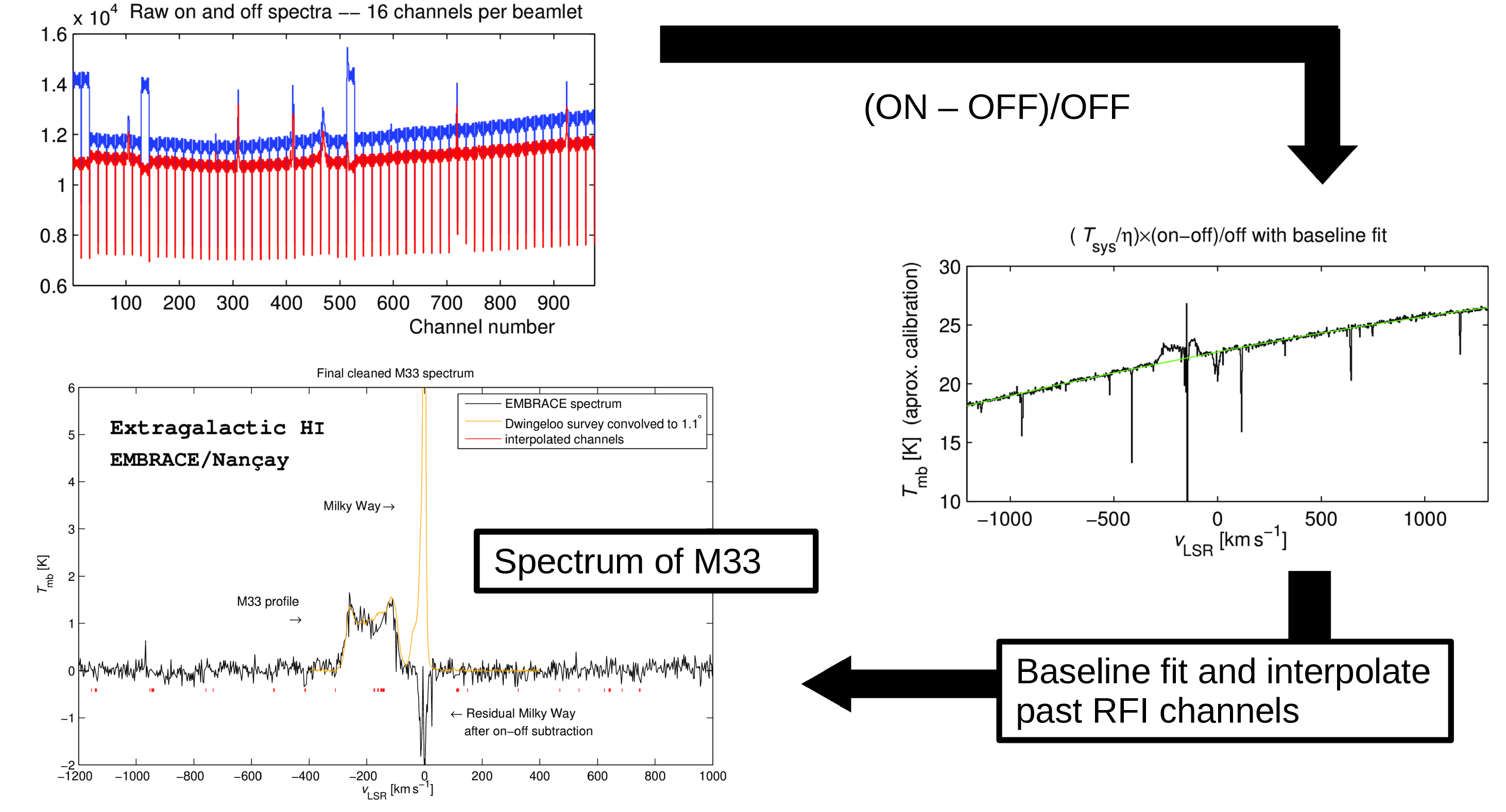}
\caption{\textbf{Top left:} Raw averaged spectra (blue=on, red=off).
Note that the horizontal axis has increasing frequency to the right. The
galactic foreground is visible in both positions around channel 470.
\textbf{Mid right:} Coarsely calibrated spectrum with four offset sub-bands
corrected by adjusting their level to that of neighbouring sub-bands.
Also shown in green is a second order baseline fit. \textbf{Bottom left:}
Baseline subtracted spectrum where obvious spikes have been interpolated out.
A comparison with the LAB survey spectrum is shown in golden/yellow.
The detailed line shape agreement is excellent apart from a portion that
is obviously asscociated with an unstable sub-band. Note that the galactic
foreground has mostly been subtracted out in the EMBRACE spectrum with only
a negative residual remaining.}
\label{fig:m33}
\end{figure}

The result and several steps in the process are illustrated in
Fig.~\ref{fig:m33}. This approach yields overall rather
benign baseline shapes indicating that much weaker signals can
probably be detected.   The sub-band
``staircasing'' and spikes/perturbations may be caused by
external or internal RFI.  The first portion of
Fig.~\ref{fig:m33} clearly shows that raw channelized spectra are
not dominated by thermal noise but rather a static square wave pattern
originating in the polyphase filter stage earlier in the chain (an
identical effect is seen in channelized single station LOFAR
spectra). The (on-off)/off operation appears to remove this
pattern extremely effectively.

\subsection{Pulsar \object{PSR B0329+54}}
\label{sec:psr}

\begin{figure}[ht!]
 \centering \includegraphics[width=0.9\linewidth,clip]{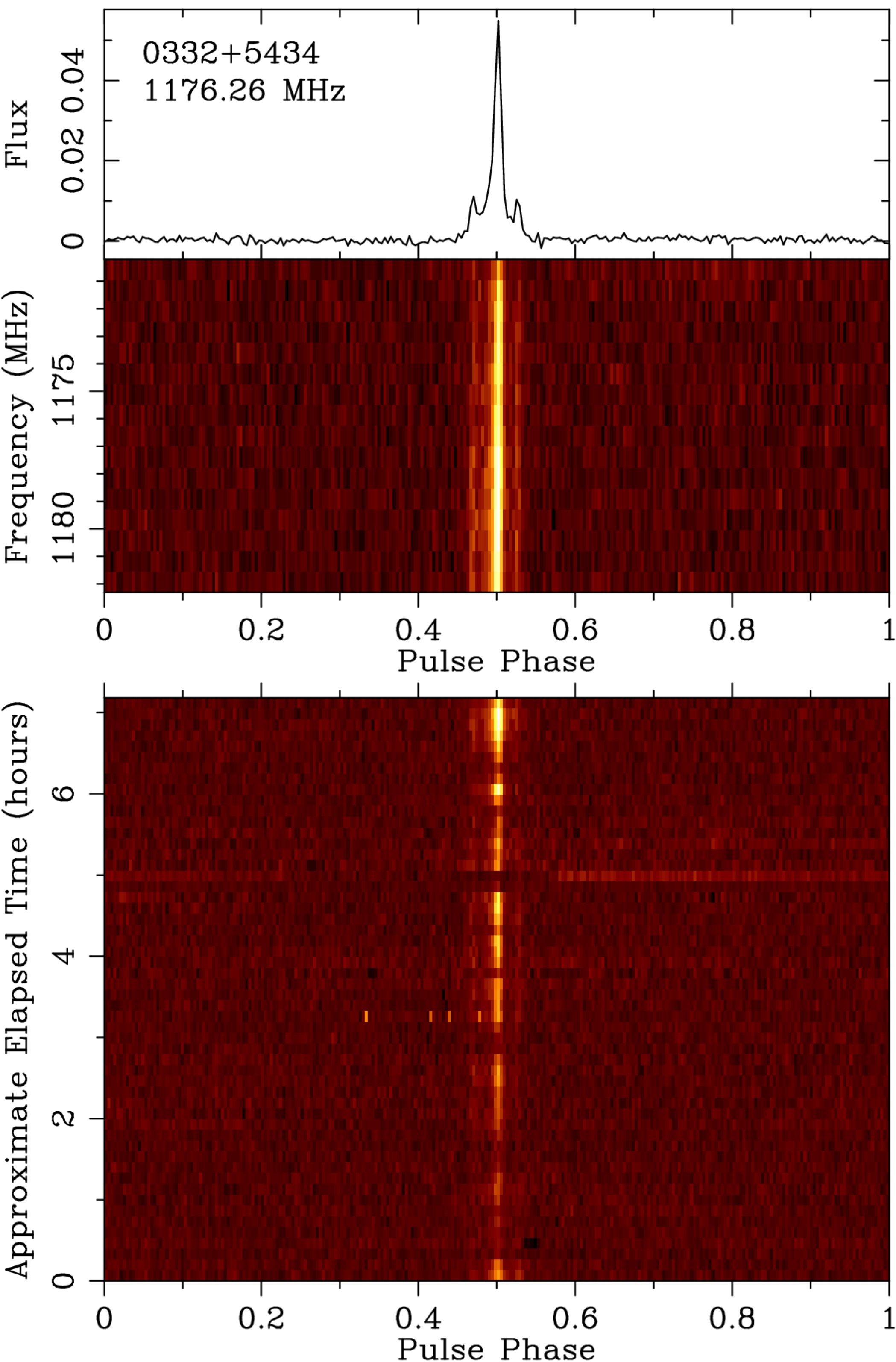}
\caption{\object{PSR B0329+54}.  The pulsar is detected after several minutes as shown in this dynamic
 plot folded at the pulsar period of 715~msec.  The figure was plotted
 using PSRCHIVE
 software~\citep{psrchive,psrchive2012}.}  \label{fig:b0329}
\end{figure}

Figure~\ref{fig:b0329} demonstrates over 9 hours of tracking
the pulsar \object{PSR B0329+54}. Its pulsed signal is clearly detected after
several minutes, and the array continues tracking, measuring
continuously the pulsar, except where RFI has been filtered at 21500
seconds. The array was configured with a bandwidth of 12~MHz (62
beamlets) centred at 1176.45~MHz. The array was phased-up using the GPS
BIIF-2 satellite, and the phase parameters were applied for the
observation of \object{PSR B0329+54}. The high data rate output from the RSP boards
is read by a data acquisition system running the Oxford ARTEMIS pulsar
processing software~\citep{2013IAUS..291..492S}.

\subsection{Daily observations}
\label{sec:dailyobs}
In November 2013, \enancay\ began a long term campaign of observation
of pulsar B0329+54.  \enancay\ has been running autonomously and
performing the pulsar observation every day centred at two
frequencies:~970~MHz and~1176~MHz (Nov~2013 to Aug~2014) and 970~MHz
and~1420~MHz (Aug~2014 to present). Since August~2014, \enancay\ has
been using saved calibration tables which greatly simplified the
observation planning.  There is no need to plan for a calibration run
before each observation.  In August 2015, drift scans of Cas~A and
Cyg~A were added to the daily observational programme.  \enancay\ is
now operating in the manner of a facility instrument, doing
pre-planned observations, and running a long term observational
campaign.

\printcomment{
{\bf insert figures: pulse profiles, driftscans}
}

\refereedelete{
\subsection{Satellites}
\label{sec:embsat}

Satellite carriers turned out to be extremely useful for various
reasons and an abundance of
observations have been done on the GPS L2 and L5 carriers (at 1227.6
and 1176.45~MHz, respectively) towards many different satellites
within the constellation.  Isolated observations of GLONASS-K1 (L3 at
1202.25~MHz) and Molniya 1-91 (downlink carrier near 991~MHz) were
also done.

Satellite carrier signals have been an important tool especially for phase
calibration procedures (see Sect.~\ref{sec:cal}).  The only other
practical option for phase calibrations turned out to be the Sun but
it does not rise above 20$^\circ$\ elevation in the middle of winter
at the telescope site and is in any case only above the formal
operational limit of 45$^\circ$\ for a few hours per day even in the
summer.  While two orders of magnitude weaker, Cassiopeia~A and
Cygnus~A are both in theory still strong enough for on-the-fly phase
calibrations but the correlator offset (see
Sect.~\ref{sec:corroffset}) imposes an additional complication to
the calibration algorithm (Sect.~\ref{sec:digcal}) which has not yet
been implemented.

\subsubsection{GPS}
GPS satellites all have orbits with inclinations of about 55$^\circ$\
and orbital periods exactly one half of a sidereal day. Since
the \enancay\ latitude is 47.4$^\circ$, many pass directly
overhead. The synchronicity with the Earth's rotation also means that
the satellites always return to the same point in the celestial sky
after two revolutions and don't have high angular rates when they
traverse the local sky.  Initial observations were tuned to the L2
carrier at 1227.6~MHz.  While useful, the signal traces still clearly
suffered from interference from neighbours due to the large number of
L2-capable satellites in the sky. The next generation of GPS
satellites were available from 2010-2011, and this new design also had
a strong broad L5 carrier at 1176.45~MHz.  At first, there were only a
handful of these GPS-BIIF type satellites and over the next three
years we had virtually no problem with signal overlap between
satellites which are well spread out on the sky at any given time. In
addition, there appear to be very few terrestrial parasitic emissions
at the location of the L5 carrier which covers $\simeq19$~MHz.

Examples of other advantages with using GPS carriers:

\vspace*{1ex}
\newlength{\itemlen}
\setlength{\itemlen}{\linewidth}
\addtolength{\itemlen}{-\parindent}
\begin{minipage}{\itemlen}
\begin{itemize}
\item Circular polarization meaning that the power received using a linear
        detector (such as EMBRACE) will be independent of relative orientations.
\item The transmitting beam is designed to provide the same gain across
        the surface of the Earth. Thus the intrinsic flux density of the
        signal does not go down (much) when a satellite is at low elevation.
\item They are true constant power point sources (whereas the Sun is 1/2 to
        1/3 of the main beam HPBW and also highly variable).
\end{itemize}
\end{minipage}

\begin{figure}
\centering
\includegraphics[width=0.9\linewidth,clip]{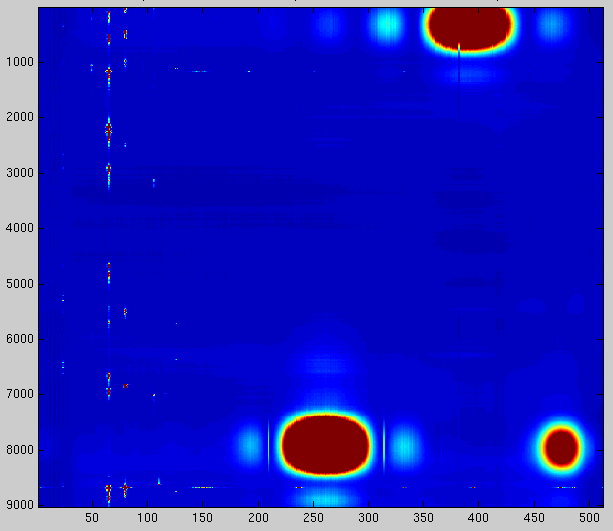}
\caption{Timeline of coaveraged tileset spectra showing how first a GLONASS
(top) and later a GPS (bottom) satellite entered the tileset beams during a
tracking observation 2012-07-13.
The $x$-axis is a sub-band count corresponding to 100~MHz of bandwidth
centred on 1176~MHz (increasing to the right), and the vertical axis is
time in seconds. The color scale is in arbitrary power units and has been
adjusted to show weaker features.}
\label{fig:gpsglo}
\end{figure}

Figure~\ref{fig:gpsglo} shows an observation of both a GPS-F and
GLONASS-K1 carrier while tracking a celestial ``empty'' sky position in order
to collect reference stability data. The satellites traversed less than
2$^\circ$\ from the centre of the tileset beams.

\subsubsection{AFRISTAR}
AFRISTAR is a geostationary satellite with subsatellite point at 21$^\circ$\
East longitude which means that it sits at a local elevation of 32.5$^\circ$\
at the \enancay\ site. It transmits radio and TV signals in the range
1469-1481~MHz, i.e., close to the upper limit of the EMBRACE band.
It was used to some extent, especially in the early testing phase,
typically for impromptu trouble shooting since it is always available
albeit at low elevation.

}

\section{Summary and future work}

The technology of dense aperture array uses a large number of antenna
elements at half wavelength spacings to fully sample the aperture.
\enancay\ is the first fully operational demonstrator of this
technology which is large enough to make interesting radio
astronomical detections.

One of the main concerns with the dense aperture array technology is
the high system complexity which could complicate operations and some
have expressed doubt that operating such a complicated instrument might not be
feasible for a facility observatory but \enancay\ has clearly
shown that dense aperture array technology is perfectly viable for
radio astronomy.
We have demonstrated its capability as a radio astronomy instrument,
including astronomical observations of pulsars and spectroscopic
observations of galaxies.  We have also demonstrated its multibeam
capability.

While the system setup is complex, once implemented, \enancay\ behaves
with remarkable stability.  System issues, such as the correlator
offset, are characterized and corrections applied to the data are
valid in the very long term and need not be remeasured frequently.
\enancay\ has been in operation for four years with little or no
change in its performance.  In particular, calibration tables used for
phase calibration continue to be valid over at least a period of
6~months.

The dense aperture array technology has the advantage of relatively
low cost fabrication being composed of large numbers of small
components which can be easily shipped and assembled on site.  After
an initial period of commissioning, the system will operate with
reliability and with predictable and stable performance.

An important disadvantage of the dense aperture array technology is
its relatively large power requirement.  The large number of analogue
electronic components associated with the signal chain of each antenna
element, together with the digital processing requirements for
beamforming and/or aperture synthesis, makes a system which is rather
power hungry.  A number of solutions are being studied to reduce the
overall power consumption~\citep{vaate2014}, including the use of high
speed digital samplers which will eliminate the frequency
down-coversion stage and not only reduce overall power consumption but
will also solve the problem of correlator offset due to a distributed
Local Oscillator signal.

Dense aperture array technology is a viable solution for the SKA,
offering the benefit of an enormous field of view together with
great flexibility for system setups, including multibeaming with
multiple and independent observation modes.  The SKA built using dense
aperture array technology will be the most rapid astronomical survey
machine.

\begin{acknowledgements}
  EMBRACE was supported by the European Community Framework Programme
  6, Square Kilometre Array Design Studies (SKADS), contract no
  011938.  We are grateful to ASTRON for initiating and developing the
  EMBRACE architecture.  MS acknowledges the financial assistance of
  the South African SKA Project (SKA SA).  AOHO, AK, and MS, were
  supported for multiple working visits to \nancay\ by grants from the
  Scientific Council of the Paris Observatory.
\end{acknowledgements}

\newcommand{\limelettebib}[3]{
  ``#1'', 
2009 {\it Proc. Wide Field Science and Technology for the SKA},
Limelette, Belgium, 
S.A.~Torchinsky \etal\ (eds), PoS(SKADS~2009)#2, {\tt 2009wska.confE..#3}}

\end{document}